\documentclass[12pt]{article}

\usepackage{fullpage}
\usepackage{color}
\usepackage{latexsym}
\usepackage{mathrsfs}
\usepackage{amsmath,amssymb}
\usepackage{bbm} 

\usepackage{multirow} 

\makeatletter \@addtoreset{equation}{section}
\makeatother

\newcommand{\be}{\begin{equation}}
\newcommand{\ee}{\end{equation}}

\newcommand{\bea}{\begin{eqnarray}}
\newcommand{\eea}{\end{eqnarray}}
\newcommand{\nn}{\nonumber}

\newcommand{\p}{\partial}
\newcommand{\ti}{\widetilde}

\newcommand{\diff}{\mathrm{d}}

\newcommand{\half}{\frac{1}{2}}

\newcommand{\lp}{\left(}
\newcommand{\rp}{\right)}

\newcommand{\bZ}{\ensuremath{\mathbb{Z}}}

\newcommand{\scA}{\ensuremath{\mathcal{A}}}

\newcommand{\scD}{\ensuremath{\mathcal{D}}}
\newcommand{\scE}{\ensuremath{\mathcal{E}}}

\newcommand{\scL}{\ensuremath{\mathcal{L}}}
\newcommand{\scM}{\ensuremath{\mathcal{M}}}
\newcommand{\scN}{\ensuremath{\mathcal{N}}}

\newcommand{\scS}{\ensuremath{\mathcal{S}}}

\newcommand{\scV}{\ensuremath{\mathcal{V}}}
\newcommand{\scW}{\ensuremath{\mathcal{W}}}

\newcommand{\Fcs}{G}

\begin{document}

\begin{titlepage}

\begin{center}

\begin{flushright}
KCL-MTH-14-18
\end{flushright}

\today

\vskip 2.5 cm 

{\LARGE \bf Supersymmetric counterterms\\ [3mm] from new minimal supergravity}

\vskip 1cm

{Benjamin Assel$^1$, Davide Cassani$^{2}$ and Dario Martelli$^1$}

\vskip 1cm

{\textit{$^1$ Department of Mathematics, King's College London, \\ [1mm]
The Strand, London WC2R 2LS,  United Kingdom}\\ [1mm]
{\small \texttt{benjamin.assel, dario.martelli AT kcl.ac.uk}}\\ [5mm]
\textit{$^2$ Sorbonne Universit\'es UPMC Paris 06,\\ [1mm] 
UMR 7589, LPTHE, F-75005, Paris, France\\ [1mm]
and\\ [1mm]
CNRS, UMR 7589, LPTHE, F-75005, Paris, France\\ [1mm]}
{\small \texttt{davide.cassani AT lpthe.jussieu.fr}}
}

\vskip 6mm

\end{center}

\vskip 1.2 cm

\begin{center}
{\bf Abstract}
\end{center}
\begin{quote}

We present a systematic classification of counterterms of four-dimensional supersymmetric field theories on curved space, obtained as the rigid limit of new minimal supergravity. These are supergravity invariants constructed using the field theory background fields. We demonstrate that if the background preserves two supercharges of opposite chirality, then all dimensionless counterterms vanish, implying that in this case the supersymmetric partition function is free of ambiguities. When only one Euclidean supercharge is preserved, we describe the ambiguities that appear in the partition function, in particular in the dependence on marginal couplings. 

\end{quote}

\end{titlepage}

\pagestyle{plain}
\setcounter{page}{1}
\newcounter{bean}
\baselineskip 6 mm

\tableofcontents

\section{Introduction and summary}

The complete information about a quantum field theory is encoded in 
 the generating functional of correlation functions of gauge-invariant operators, 
 defined as the Euclidean path integral over the dynamical fields of the theory in the presence of background sources. 
Although this is usually very difficult to compute in closed form,  for certain supersymmetric field theories defined on compact Riemannian manifolds 
supersymmetric localization can be used to simplify the calculation enormously \cite{Witten:1988ze,Nekrasov:2002qd,Pestun:2007rz}. 
As recently emphasised in~\cite{Festuccia:2011ws}, supersymmetric field theories on curved manifolds can be constructed by taking 
a rigid limit of a suitable off-shell supergravity theory, where gravity is decoupled by sending the Planck mass to infinity.
In this limit, the metric and other auxiliary 
fields in the supergravity multiplet  remain as  background fields and play the role of sources in the functional integral.  

Although localization techniques allow to reduce the path integral to a semi-classical calculation of a one-loop functional determinant, the latter is a divergent quantity, that needs to be regularised and renormalised.  Moreover, the finite part might suffer from ambiguities associated to the different renormalisation schemes, as usual in quantum field theories. Divergences can be removed by adding  \emph{counterterms}, namely  integrals of local densities constructed from the background fields.
Ambiguities in the finite part of the calculations  correspond to the existence of counterterms that remain finite after removing the cut-off.
The choice of renormalisation scheme, and therefore of the possible counterterms, is constrained by the requirement that certain symmetries of the theory be preserved. 

Motivated by the need to characterise physically meaningful observables,
in this paper we classify counterterms of four-dimensional $\scN=1$ supersymmetric field theories on curved space. In this case, the relevant counterterms are supergravity invariants. One way to see it is to observe that some higher-derivative terms in the off-shell supergravity action do not involve the Planck mass, and therefore survive the rigid limit. 
 From the field theory viewpoint, these dimensionless integrals involving just background fields define finite supersymmetric counterterms. Supergravity invariants involving positive powers of the Planck mass provide counterterms that may be used to subtract UV divergences.

The form of the counterterms depends on the specific off-shell supergravity one starts from. Here we consider the new minimal formulation of supergravity~\cite{Sohnius:1981tp,Sohnius:1982fw}, which assumes the existence of an R-symmetry. As the field theory path integral is usually defined on Riemannian manifolds, we will work in Euclidean signature. 
 The first part of our work is thus  devoted to define new minimal supergravity in Euclidean signature. This is achieved by doubling the number of degrees of freedom of all fields. Supersymmetric invariants are then constructed either as $D$-terms, or $F$-terms, or $\ti F$-terms (in Lorentzian signature, $\ti F$-terms are the complex conjugate to the $F$-terms, but in Euclidean signature they are independent) In the three-dimensional avatar of new minimal supergravity, local supersymmetric invariants corresponding to finite counterterms were studied in \cite{Closset:2012vp,Closset:2012vg}. Supersymmetric invariants of other supergravity theories  were recently used as counterterms in \cite{Gerchkovitz:2014gta,Gomis:2014woa}.\footnote{In Lorentzian signature, supersymmetric invariants of new minimal supergravity have been 
 discussed in \cite{Brandt:1996au} in the context of local BRST cohomology. We thank the author for alerting us about the relevance of this work.}

We first analyse the terms that are constructed  exclusively with component fields of  the gravity multiplet. In four-dimensional new minimal supergravity, the bosonic fields are the vielbein $e^a_\mu$ and two auxiliary fields: a one-form $A_\mu$ gauging the R-symmetry and a conserved, globally defined one-form $V_\mu$. Besides locality and supersymmetry, we also demand that the counterterms be invariant under diffeomorphisms and local 
R-symmetry transformations. This selects a set of curvature multiplets as building blocks, which can be combined by exploiting the tensor calculus of new minimal supergravity~\cite{Sohnius:1982fw,Ferrara:1988qxa}.

We will call {\it marginal} the action obtained by integrating a local Lagrangian of mass dimension $d=4$ over the four-manifold. Having vanishing mass dimension, this naturally leads to a finite counterterm (the same type of integral can also multiply logarithmic divergences). On the other hand, we will call {\it dimensionful} the integral of a local Lagrangian of mass dimension $d < 4$. To form an action, this must be multiplied by $\Lambda^{4-d}$, where $\Lambda$ is a parameter with the dimension of a mass. This may be a parameter appearing in the field theory Lagrangian, or a UV cut-off scale in a given regularisation scheme. In the second case, dimensionful integrals provide divergent counterterms.

We find that a basis of independent marginal supersymmetric actions is given by\footnote{Here we only write  the bosonic parts.  Precise definitions of $\mathscr E$ and $\mathscr P$ are given in~\eqref{def_EulerDensity} and~\eqref{def_PontryDensity}.}
\bea
S_{\mathscr E} &=&  \int \diff^4 x \, e\, \mathscr E \ = \ 32\pi^2 \, \chi (M) \ ,\nn \\ [2mm]
S_{\rm \mathscr P} &=& \int \diff^4 x \, e \,\mathscr P \  = \  48\pi^2 \,\sigma  (M) \ ,\nn \\ [2mm]
S_F &=& \int \diff^4 x \,e\,  F_{\mu\nu} *\!F^{\mu\nu}  \ = \ 2\pi^2\, \nu\ , \label{FiniteCounters} \\ [2mm]
S_{C^2} &=& \int \diff^4 x \, e  \, \big[ C_{\mu\nu\rho\sigma}C^{\mu\nu\rho\sigma} - \frac{8}{3} \Fcs_{\mu\nu} \Fcs^{\mu\nu} \big]\ ,\nn \\ [2mm]
S_{R^2} &=&  \int \diff^4 x \, e \, \big[8 F_{\mu\nu} F^{\mu\nu} - \left( R + 6 V_\mu V^\mu\right)^2 \big]\ , \nn
\eea
where $e$ is the vielbein determinant, $\mathscr E$ denotes  the Euler density, integrating to the Euler characteristic $\chi (M)$ of the four-manifold $M$, $\mathscr P$ denotes the Pontryagin density, integrating to the signature $\sigma (M)$, 
$F_{\mu\nu} $ is the field strength of $A_\mu$, $C_{\mu\nu\rho\sigma}$ is the Weyl tensor of the Levi-Civita connection, $G_{\mu\nu}$ is the field strength of $A_\mu - \frac 32 V_\mu$, and $R$ is the Ricci scalar of the Levi-Civita connection. The first three integrals in (\ref{FiniteCounters}) are topological invariants, the fourth is conformal invariant, while the fifth is neither topological, nor conformal. Let us also recall that when these integrals are evaluated on a generic supersymmetric background, the  instanton number $\nu$ gets related to the topological invariants of the manifold via the formula $\nu = c_1(M)^2=3 \sigma (M)+2 \chi (M)$ \cite{Cassani:2013dba}.
In Lorentzian signature, these invariants were constructed in~\cite{Cecotti:1987qe,Ferrara:1988qxa,deRoo:1990zm}, and  our study indicates that there exist no others satisfying the requirements.

Regarding dimensionful integrals, it is known that there is no supersymmetrisation of the cosmological constant in new minimal supergravity~\cite{Sohnius:1981tp,Sohnius:1982fw}, and we interpret this as a proof that quartic divergences in the UV cut-off cannot appear if the field theory is regularised in a supersymmetric way.
The only dimensionful integral we find is the supergravity Einstein--Hilbert (EH) action, whose bosonic part reads~\cite{Sohnius:1981tp}
\be\label{EHaction}
S_{\rm EH} \ =\ \frac{\Lambda^2}{2} \int \diff^4 x \, e \left(R + 6 V_\mu V^\mu - 8 A_\mu V^\mu \right)\ .
\ee
Therefore this is the only supersymmetric counterterm that could be used to subtract quadratic divergences, if indeed these arise.

We can also consider further background multiplets in addition to the gravity multiplet. For instance, one can introduce background gauge vector multiplets, coupling to field theory flavor supercurrents. The associated Yang--Mills and topological actions naturally provide marginal supersymmetric counterterms, while Fayet--Iliopoulos (FI) terms of background Abelian gauge vector multiplets are dimensionful and come multiplied by $\Lambda^2$, as \eqref{EHaction}. By combining the fields in the gravity multiplet and in the background gauge vector multiplet, one can also construct mixed gravity-gauge counterterms whose expression we will present.

Further supersymmetric counterterms can be defined by promoting the couplings appearing in the field theory Lagrangian to background multiplets, and using these to construct new supersymmetric actions \cite{Seiberg:1993vc} (see also \emph{e.g.} \cite{Dine:1996ui,Weinberg:2000cr}). As emphasised in \cite{Gerchkovitz:2014gta}, these counterterms parameterise the ambiguous dependence of the field theory partition function on such couplings.

Having constructed the supersymmetric counterterms, we proceed to evaluate them on a supersymmetric background. This is obtained by setting to zero all fermionic background fields and imposing that the respective supersymmetry variations also vanish. In particular, assuming  that the metric is real, one can show that generically the manifold is endowed with a complex structure \cite{Klare:2012gn,Dumitrescu:2012ha}.
The conclusions depend on the number of supercharges being preserved.
If the background admits two supercharges of opposite R-charge, then all supersymmetric invariants constructed as $F$-terms or $\ti F$-terms vanish. Invariants constructed as $D$-terms of a general multiplet also vanish, except if the $D$-term is an FI term of an Abelian gauge vector multiplet (as we will see, this also includes the EH action \eqref{EHaction}). However, at least if the field theory does not have a relevant parameter in the Lagrangian, such non-vanishing $D$-terms can at most define divergent counterterms. We conclude that in the presence of two supercharges of opposite R-charge, all finite, local, gauge-invariant counterterms vanish. In particular, this holds for \emph{all} terms in \eqref{FiniteCounters}. It follows that when the renormalisation scheme preserves supersymmetry as well as diffeomorphism and gauge invariance, the partition function is free of ambiguities.
 
If only one Euclidean supercharge is preserved, all $\ti F$-terms built out of background fields vanish, while $F$-terms can be non-zero (or vice-versa, depending on the chirality of the supercharge). This implies some relations between the integrals \eqref{FiniteCounters}, such that they can all be expressed in terms of topological invariants. The same holds for the marginal counterterms from background gauge vector multiplets.
Moreover, the $D$-terms behave as in the two supercharge case. It follows that, apart for an overall number fixed by the topology of the manifold and of the gauge bundle of background vector multiplets, again there are no ambiguities in the field theory partition function as far as the dependence on the background fields is concerned. However, generically we obtain non-trivial ambiguities in the dependence on marginal couplings appearing in the field theory action. It was recently shown in \cite{Closset:2014uda} that, up to counterterms and anomalies, the partition function is a locally holomorphic function of $F$-terms couplings. By promoting the marginal couplings to background chiral multiplets, we will show that as long as at least one of the topological 
invariants mentioned above does not vanish, there is an ambiguity by an arbitrary holomorphic function of such couplings.
 
In addition, the present results extend those found in~\cite{Cassani:2013dba}, where the first four integrals in \eqref{FiniteCounters} were evaluated on a supersymmetric background, but no analysis was performed of other possible terms. The terms studied in~\cite{Cassani:2013dba} are those appearing in the Weyl anomaly and chiral anomaly of the R-current of superconformal field theories, namely
\bea 
\int \diff^4 x \, e\, \langle T_\mu{}^\mu \rangle & = &  \frac{c}{16\pi^2} S_{C^2} - \frac{a}{16\pi^2}  S_{\mathscr E}~, \nn\\ [1mm]
\quad\; \int \diff^4 x \, e\, \langle \nabla_\mu J_{\rm R}^\mu \rangle & = &  \frac{c-a}{24\pi^2}  S_{\mathscr P} + \frac{5a-3c}{27\pi^2} S_G \ ,
\eea
where $S_G$ denotes the topological invariant $S_G = \int \diff^4 x \,e\,  G_{\mu\nu} * G^{\mu\nu}$ which, although {\it a priori} independent, is equal to $S_F$ since $V_\mu$ is globally defined.
In particular, it was observed in~\cite{Cassani:2013dba} that when two supercharges of opposite R-charge are preserved, the integrated trace of the energy-momentum tensor and the divergence of the R-current  vanish. Here we will show that it is immediate to extend these results to the presence of background gauge vector multiplets and to chiral anomalies of flavor currents.

Finally, let us briefly comment on the same problem studied in Lorentzian signature. Assuming a suitable fall-off of all fields at infinity such that surface terms from integrations by parts evaluate to zero, the picture is exactly the same as the one for the case of two Euclidean supercharges of opposite R-charge (there is no analog of the one Euclidean supercharge case in Lorentzian signature). See~\cite{Cassani:2012ri} for a characterisation of Lorentzian supersymmetric backgrounds in four dimensions.

The rest of the paper is organised as follows. In section~\ref{SusyTransfoSection} we discuss new minimal supergravity in Euclidean signature. In section~\ref{sec:ConstrCount} we construct the supersymmetric counterterms. In section~\ref{sec:VanishingResults} we evaluate them on a supersymmetric background, and prove our vanishing results. We then discuss some physical implications: in section~\ref{sec:PhysicalImplications} regarding ambiguities and in section~\ref{sec:anomalies} about anomalies. Section~\ref{sec:conclusions} contains our concluding remarks. Appendix~\ref{app:conventions} provides our conventions and various useful identities, while appendix~\ref{app:TensorCalculus} gives more details on the tensor calculus.
In Appendix~\ref{app:CompareOldMinimal} we make some comments on the differences with old minimal supergravity.

\section{Euclidean new minimal supergravity}\label{SusyTransfoSection}

In this section we review the relevant features of new minimal supergravity \cite{Sohnius:1981tp, Sohnius:1982fw}. We mainly follow ref.~\cite{Ferrara:1988qxa}, which includes a thorough study of the tensor calculus and of the curvature multiplets. Having in mind applications to field theories defined on Riemannian manifolds, we work in Euclidean signature. We start introducing the gravity multiplet and its transformation rules. Then we  discuss the general multiplet, the different types of special multiplets (chiral, gauge, linear) and the curvature multiplets, providing just the essential notions that will be needed in the next sections.
Besides Wick-rotating the Lorentz and Clifford algebras, the Euclidean supergravity theory has been obtained from the Lorentzian one by splitting all Majorana spinors in their positive and negative chirality parts, and allowing these to be independent of each other. It follows that the supersymmetry transformations do not preserve any reality property, so we should allow all bosonic fields, including the metric, to generically take complex values. This effectively doubles the number of all off-shell degrees of freedom.

\subsection{Gravity multiplet}

The gravity multiplet of Euclidean new minimal supergravity is
\be
\big (e^a_\mu\,,\, \psi_{\mu\alpha}\,,\, \ti\psi_{\mu}^{\dot\alpha}\,,\, A_{\mu}\,,\, B_{\mu\nu} \big)\,,
\ee
where $e^a_\mu$ is the vierbein, $\psi_{\mu\alpha}$, $\ti \psi_\mu^{\dot\alpha}$ are independent gravitini of positive and negative chirality, respectively (see appendix~\ref{app:conventions} for details about our spinor conventions), while $A_\mu$ and $B_{\mu\nu}$ are auxiliary fields. $A_\mu$ is the gauge field of a local, Abelian R-symmetry, under which $\psi_\mu$ has charge $+1$ and $\ti\psi_{\mu}$ has charge $-1\,$. $B_{\mu\nu}$ is an antisymmetric tensor with gauge transformation $\delta B_{\mu\nu} = 2 \partial_{[\mu}\xi_{\nu]}$. Since all other fields (including those in the matter multiplets to be introduced later) are neutral under the latter transformation, $B_{\mu\nu}$ can equivalently be described in terms of a one-form $V_\mu$, defined as the Hodge dual of the field strength $H_{\mu\nu\rho} = 3 \,\partial_{[\mu}B_{\nu\rho]}\,$:
\be\label{HdualV}
V_\mu \ =\ - \frac i 6 \epsilon_{\mu\nu\rho\sigma} H^{\nu\rho\sigma}\,.
\ee
The Bianchi identity $\partial_{[\mu}H_{\nu\rho\sigma]}=0$ translates into the conservation condition $\nabla_\mu V^\mu = 0\,$. We will find convenient to use both $V_\mu$ and $H_{\mu\nu\rho}$ in the following.

The supersymmetry transformations are defined in terms of spinorial parameters $\zeta$ and $\ti\zeta$, having the same chirality and R-charge as $\psi_\mu$ and $\ti \psi_\mu\,$, respectively. Before coming to their expression, it is convenient to include suitable gravitino bilinears into the definition of various bosonic quantities. This repackages the complicated gravitino terms of supergravity, allowing to handle them in a relatively simple way, and making the expressions supercovariant.\footnote{This implies that the supersymmetry variation of hatted quantities does not contain derivatives of spinorial parameters.} Such modified bosonic quantities will be denoted by a hat symbol.
As a first thing, as usual in supergravity, we define a spin connection $\hat \omega_\mu{}^{ab}$ with torsion, where the torsion tensor is $T_{\mu\nu}{}^a = - 4 i   \ti \psi_{[\mu}\ti\sigma^a \psi_{\nu]}\,$.\footnote{Explicitly, $\hat \omega_\mu{}^{ab}  =  \omega_{\mu}{}^{ab} + K_\mu{}^{ab}$, where $\omega_\mu{}^{ab}$ is the Levi-Civita connection defined in \eqref{def_spin_conn}, and the contortion is $K_\mu{}^{ab} \, =\, -2i \big(\ti \psi_\mu \ti \sigma^{[a} \psi^{b]} + \ti\psi^{[a} \ti \sigma^{b]} \psi_\mu +  \ti\psi^{[a} \ti \sigma_\mu \psi^{b]} \big)\,.$}
We will further define 
\bea
\hat H_{\mu\nu\rho} &=& H_{\mu\nu\rho} + 6 i \,  \psi_{[\mu} \sigma_{\nu} \ti \psi_{\rho]}\ ,\nn \\ [2mm]
\hat V_{\mu} &=& V_{\mu} +   \,\epsilon_\mu{}^{\nu\rho\sigma} \, \psi_\nu \sigma_\rho\ti \psi_\sigma \ .
\eea
Then we can introduce the connections
\be
\omega^\pm_{\mu}{}^{ab} \ =\ \hat \omega_\mu{}^{ab} \pm \hat H_{\mu}{}^{ab} \ = \  \hat \omega_\mu{}^{ab} \pm i \, \hat V^\nu\epsilon_{\nu\mu}{}^{ab}\ ,
\ee
whose torsion also includes a bosonic part proportional to $H_{\mu ab}$, and
\be
A^+_\mu \ =\ A_\mu - \hat V_\mu\ ,
\ee
the notation being inherited from~\cite{Ferrara:1988qxa}. These can be used to define covariant \hbox{derivatives $D^\pm$:}
\bea
D^{+}_\mu &=&  \partial_\mu + \frac i2\omega^+_\mu{}^{ab}S_{ab}  - i r A^+_\mu \nn \\ [2mm]
D^{-}_\mu &=& \partial_\mu + \frac i2\omega^-_\mu{}^{ab}S_{ab}  - i r A_\mu \ .
\eea
Here, $S_{ab}$ denotes the (Hermitian) generators of the $SO(4)$ Lorentz rotations, acting on a field $\Phi$  as $\delta_{\rm Lorentz}(\lambda) \Phi = \frac i2 \lambda^{ab} S_{ab} \Phi$, while $r$ is the R-charge of the field that is acted on.  For instance, for the gravitino $\psi_\mu$, $S_{ab} = i \sigma_{ab}$ and $r=1$, while for $\ti\psi_\mu$,  $S_{ab}= i\, \ti \sigma_{ab}\,$ and $r=-1$. On the supersymmetry parameters, the derivative $D^+$ gives
\bea
D_\mu^+ \zeta \!&=&\! \left(\partial_\mu - \tfrac 12 \omega^+_{\mu ab}\sigma^{ab} - i A^+_\mu \right) \zeta \; = \;  \big( \partial_\mu - \tfrac 12 \hat\omega_{\mu ab}\sigma^{ab} - i A_\mu + i \hat V_{\mu}  + i \hat V^{\nu} \sigma_{\mu\nu}\big) \zeta\ ,\nn \\ [2mm] 
D_{\mu}^+ \ti \zeta \!&=&\! \left(\partial_\mu - \tfrac 12 \omega^+_{\mu ab}\ti \sigma^{ab} + i A^+_\mu \right) \ti\zeta \; = \; \big( \partial_\mu - \tfrac 12 \hat\omega_{\mu ab}\ti\sigma^{ab} + i A_\mu - i \hat V_{\mu}  - i \hat V^{\nu}\, \ti\sigma_{\mu\nu}\big) \ti\zeta \ .\quad\label{D+SpinorParam}
\eea

We will also need the gravitino field strengths, defined as
\be
\psi_{\mu\nu} \ = \  D^+_{\mu} \psi_{\nu} - D^+_{\nu} \psi_{\mu}
\,,\qquad \ti \psi_{\mu\nu} \ = \  D^+_{\mu} \ti \psi_{\nu} - D^+_{\nu} \ti \psi_{\mu}\ .
\ee

We can now give the supersymmetry transformations of the gravity multiplet. These are:
\bea
\delta e^a_{\mu} &=&   2 i\, \zeta \sigma^{a} \ti\psi_{\mu} + 2 i\, \ti \zeta\, \ti\sigma^{a} \psi_{\mu} \ , \nn\\ [2mm]
\delta \psi_{\mu} &=& D_\mu^+ \zeta\ , \nn\\ [2mm]
\delta \ti\psi_{\mu} &=&  D_\mu^+ \ti\zeta\ , \nn\\ [2mm]
\delta A_{\mu} &=&      \zeta \sigma_\mu \ti \sigma^{ab} \ti \psi_{ab}  -   \ti \zeta \,\ti \sigma_\mu  \sigma^{ab} \psi_{ab}  \ ,  \nn\\ [2mm]
\delta B_{\mu\nu} &=& -   \, 2 i\, \zeta \sigma_{[\mu} \ti\psi_{\nu]}  - 2 i\,\ti\zeta\, \ti\sigma_{[\mu} \psi_{\nu]} \ .\label{SusyVarGravity}
\eea
We also provide a few more supersymmetry variations, which can be deduced from the ones above and will be useful in the following.
The variation of the spin connection is
\bea\label{VarHatOmega}
\delta \hat \omega_{\mu ab} &=& i   \left(- \zeta \sigma_\mu \ti\psi_{ab} + \zeta \sigma_a \ti\psi_{b \mu} - \zeta \sigma_b \, \ti\psi_{a \mu} + 2 \hat H_{abc} \, \zeta \sigma^c \ti\psi_\mu \right)\nn \\ [2mm]
&&\!\!\!\!\!\!+\, i   \left(- \ti \zeta \, \ti \sigma_\mu \psi_{ab} + \ti \zeta \,\ti \sigma_a \psi_{b \mu} - \ti \zeta \,\ti \sigma_b \, \psi_{a \mu} + 2 \hat H_{abc} \,\ti \zeta \,\ti\sigma^c \psi_\mu \right)\,.
\eea
The variation of $\hat H_{abc}$ is
\be\label{VarHatHflat}
\delta \hat H_{abc} \ =\ 3i  \big(\zeta   \sigma_{[a} \ti\psi_{bc]} +  \ti\zeta\,\ti\sigma_{[a} \psi_{bc]}  \big) \ .
\ee
Equivalently,
\be
\delta  \hat V_a \ = \ \frac 12   \epsilon_a{}^{bcd} \, \big( \zeta   \sigma_b \ti\psi_{cd} + \ti \zeta \, \ti \sigma_b \psi_{cd} \big)  \ .
\ee
The variations of $\omega_{\mu ab}^{\pm}$ are
\bea
\delta \omega^+_{\mu ab} &=&  4 i   \big( \zeta \sigma_{[a} \ti\psi_{b] \mu}+ \ti\zeta \,\ti \sigma_{[a} \psi_{b] \mu}\big)  + 4 i  \hat H_{abc} \,\big( \zeta \sigma^c \ti \psi_\mu  + \ti \zeta \,\ti \sigma^c \psi_\mu \big)\ ,\\ [2mm]
\delta \omega^-_{\mu ab} &=& - 2i   \big( \zeta \sigma_\mu \ti \psi_{ab} + \ti \zeta\,\ti \sigma_\mu \psi_{ab}\big)\ .\label{VarOmegaMinus}
\eea
Note that the second is particularly simple. 
It will also be useful to record that
\be
\delta A^+_\mu \ = \  - \zeta \sigma^a \ti\psi_{\mu a} + \ti \zeta\,\ti \sigma^a \psi_{\mu a}  - 2 i \hat V_a \big( \zeta \sigma^a \ti\psi_\mu + \ti \zeta\,\ti \sigma^a \psi_\mu \big) \ .
\ee
The variation of the gravitino field strengths can be expressed as
\be\label{VarTildePsiab_bis}
\delta  \psi_{ab} \ = \ -\frac 12 \hat R^+_{abcd}\sigma^{cd}\zeta - i \hat F^+_{ab} \, \zeta\ , \qquad
\delta \ti \psi_{ab} \ = \ -\frac 12 \hat R^+_{abcd}\ti\sigma^{cd}\ti \zeta + i \hat F^+_{ab} \,\ti \zeta\ .
\ee
We have defined the field strength of $A^+$ as $F_{ab}^{+}= 2\,e_a^\mu e_b^\nu \partial_{[\mu} A^+_{\nu]}$. Moreover, $R^+_{abcd}= e_a^\mu e_b^\nu R^+_{\mu\nu cd}$ is the Riemann tensor computed from the connection with torsion $\omega^+_{\mu ab}$; its expression at vanishing gravitino is given in eq.~\eqref{curvature_Htorsion}. The hatted curvatures are the truly supercovariant curvatures, obtained by using a modified covariant derivative 
\be
\hat D^\pm_\mu \ =\ D^\pm_\mu - \delta_Q(\psi_\mu,\ti\psi_\mu)\ ,
\ee 
where the last term denotes a supersymmetry variation with parameters $\psi_\mu$, $\ti \psi_\mu$ at the place of $\zeta$, $\ti\zeta\,$.

\subsection{General multiplet and subcases}

We will now introduce the general multiplet of new minimal supergravity and some of its sub-multiplets \cite{Sohnius:1982fw,Ferrara:1988qxa}. As we work in Euclidean signature, we adopt the notation of \cite[sect.\:2]{Closset:2014uda} (extended to supergravity).
A general supermultiplet $\scS$ has components
\be
\scS \ = \ \big(C\,,\, \chi_\alpha \,,\, \ti\chi{}^{\dot \alpha} \,,\, M \,,\,\ti M \,,\, a_{\mu} \,,\, \lambda_\alpha \,,\, \ti\lambda^{\dot\alpha} \,,\, D \big)\ .
\ee
Its R-charge $r$ is defined as the R-charge of the bottom component $C$, and the R-charges of the different components are $(r,r-1,r+1,r-2,r+2,r,r+1,r-1,r)$. The supersymmetry transformations are:
\bea
\delta C &=& i \zeta\chi - i \ti\zeta\, \ti\chi \ ,\nn\\ [2mm]
\delta \chi &=& \zeta M + i\, \sigma^{b} \ti\zeta \big(   a_{b} -i  \hat D^-_{b}C \big)\ ,
\nn\\ [2mm]
\delta \ti\chi &=& \ti\zeta \,\ti M + i\, \ti\sigma^{b} \zeta \big(   a_{b}  + i \hat D^-_{b}C \big)\ , \nn\\ [2mm]
\delta M &=& 2 \ti\zeta \, \ti\lambda + 2 i\, \ti\zeta \,\ti\sigma^{a} \big( \hat D_{a}^-\chi - 2i\hat V_a\, \chi  \big)  - 2i \, \ti\zeta\, \ti\Xi\, C\ ,
\nn\\ [2mm]
\delta \ti M &=& 2 \zeta \lambda + 2 i \,\zeta \sigma^{a} \big( \hat D_{a}^- \ti\chi + 2i\hat V_a\, \ti\chi  \big) + 2i \,\zeta\, \Xi\, C\ ,
\nn\\ [2mm]
\delta a_{b} &=& i \big( \zeta\sigma_{b}\ti\lambda + \ti\zeta\, \ti\sigma_{b}\lambda \big) 
+  \zeta \big(\hat D^-_{b}\chi + i V^a \sigma_b \ti\sigma_a \chi   \big)  + \ti\zeta \big( \hat D^-_{b}\ti\chi - i V^a \ti\sigma_b \sigma_a \ti\chi \big)\ ,  \nn \\ [2mm]
\delta \lambda &=&    i \zeta D + 2 \sigma^{ab} \zeta \Big( \hat D^-_{a} a_{b}  +2i \hat V_a a_b - \frac 12 \psi_{ab} \chi - \frac 12 \ti\psi_{ab} \ti\chi \Big) 
+  \Xi (\zeta\chi + \ti\zeta \ti\chi)\ ,  \nn\\ [2mm]
 \delta \ti\lambda &=&   - i \ti\zeta D + 2 \ti\sigma^{ab}\ti\zeta  \Big( \hat D^-_{a} a_{b}  - 2i \hat V_a a_b - \frac 12 \psi_{ab} \chi - \frac 12 \ti\psi_{ab} \ti\chi \Big) 
+ \ti\Xi (\zeta\chi + \ti\zeta \ti\chi)\ , \nn\\ [2mm]
\delta D &=& -   \zeta\sigma^{b}  \lp \hat D^-_{b} \ti\lambda  - \ti\Xi\, a_b \rp 
+ \ti\zeta\, \ti\sigma^{b}  \lp \hat D^-_{b}\lambda - \Xi\, a_b \rp   
  +   (\zeta\, \Delta \chi + \ti\zeta\, \Delta\ti\chi\,) 
 \ ,
\label{GenMultTransfo}
\eea
where
$\Xi = - i\psi^{ab} (S_{ab} +  i \sigma_{ab}\, r) $, $\ti\Xi = -i\ti\psi^{ab}( S_{ab} - i\,  \ti \sigma_{ab}\, r)$ and $\Delta = i \hat F^+_{ab} S^{ab} - \frac{i}{4}\hat R^+ r $.

Two general multiplets can be multiplied using the rules of tensor calculus \cite{Sohnius:1982fw,Ferrara:1988qxa}. These are reviewed (at vanishing gravitino) in appendix \ref{MultipletProduct}.

We now briefly present some special multiplets obtained imposing the vanishing of at least one of the components of the general multiplet.

\medskip

{\bf Chiral multiplet.} This is defined by the condition  $\widetilde \chi{}^{\dot\alpha} = 0$, leading to a multiplet $\Phi$ with independent components $\Phi = (\phi,\,\psi_\alpha,\,F)$. It is embedded in a general multiplet of R-charge $r$ as
\be
\Phi \ =\ \left(\,\phi\,,\, -\sqrt 2i\psi_\alpha\,,\,0\,,\,-2iF\,,\,0\,,\,-i \hat D^-_b \phi\,,\,  \ldots ,\, i \Delta \phi\, \right)\ ,
\ee
where ``$\ldots$'' denote terms involving gravitino fields, that we will not need.

\medskip

{\bf Anti-Chiral multiplet.} It is defined by the condition $\chi_\alpha=0$, leading to a multiplet $\ti \Phi$ with independent components $\ti \Phi = \big(  \ti \phi ,\, \ti \psi^{\dot \alpha},\,\ti F\big)$. It is embedded in a general multiplet of R-charge $r$ as
\be
\ti \Phi \ = \ \left( \,\ti \phi\,,\,0\,,\, \sqrt 2i\, \ti\psi^{\dot \alpha}\,,\,0\,,\,2i\ti F\,,\, i \hat D^-_b \ti\phi\,,\, \ldots \,,\, -i \Delta \ti \phi\, \right)\,.
\ee
In Lorentzian signature, $\ti \Phi$ of R-charge $r=-q$ is the conjugate of $\Phi$ with $r=q$.

\medskip

{\bf Gauge vector multiplet.} A gauge vector multiplet in Wess--Zumino gauge, ($a_{\mu}, \lambda_\alpha, \ti\lambda^{\dot \alpha}, D$), has $C=M=\ti M= \chi = \ti\chi =0$. From this, one can construct field strength multiplets $\Lambda_\alpha$ and $\ti \Lambda^{\dot \alpha}$, which carry a fermionic index and are chiral and anti-chiral multiplets, respectively. $\Lambda_\alpha$ is defined as the multiplet whose bottom component is the positive-chirality gaugino $\lambda_\alpha$ of the gauge vector multiplet, while $\ti\Lambda^{\dot\alpha}$ as the multiplet whose bottom component is the negative-chirality gaugino $\ti\lambda^{\dot\alpha}$. We give explicit formulae in appendix~\ref{FieldStrength}.

\medskip

{\bf Complex linear multiplet.} It is obtained from a general multiplet by setting $\ti M = 0$. The independent components are $(C, \chi_\alpha, \ti\chi^{\dot \alpha}, M ,  a_{\mu}, \ti \lambda)$, where $a_\mu$ is a well-defined one-form.
Later we will need the $D$ component of a complex linear multiplet of vanishing R-charge and at vanishing gravitino. This reads
\be\label{Dterm_ComplexLinearMult}
D \ =\ - (D^-)^2 C + i D^{- \, b} a_b  + 2 V^b (a_b + i D^-_b C ) - i F^+_{ab} S^{ab}C\ ,
\ee
and all quantities are evaluated at $\psi_\mu = \ti \psi_\mu =0$. A similar type of complex linear multiplet is obtained by setting $M=0$.

\medskip

{\bf ``Real'' linear multiplet.} A ``real'' linear multiplet $L$ is obtained from a general multiplet by setting $M = \ti M = 0$ (the quotation marks indicate that this multiplet is truly real only in Lorentzian signature). Its independent components are $L  = (C, \chi_\alpha, \ti\chi^{\dot \alpha}, a_{\mu})$, where $a_\mu$ is a well-defined one-form. Its embedding in the general multiplet is
\be
L  \ =\ (C \,,\, \chi_\alpha\,,\, \ti\chi^{\dot \alpha}\,,\, 0\,,\, 0\,,\, a_{\mu}\,,\, \ldots ) \ .
\ee
Moreover the components are subject to a constraint which at vanishing gravitino reads 
\be 
D^{- \, b} a_b  + 2  V^b D^-_b C  -  F^+_{ab} S^{ab}C \ =\ 0\ .
\ee 
The $D$ component \eqref{Dterm_ComplexLinearMult} then reduces to 
\be\label{Dterm_RealLinearMult}
D \ = \ - (D^-)^2 \, C + 2 \, V^{b} a_b \ .
\ee

\medskip

{\bf Spinor derivatives.}
The spinor derivative operators $\scD_\alpha, \ti \scD^{\dot\alpha}$ acting on a general multiplet $\scS$ are defined by letting $\scD_\alpha \scS$ be the multiplet with bottom component $\chi_\alpha$ and $\ti \scD^{\dot\alpha} \scS$ be the multiplet with bottom component $\ti\chi^{\dot \alpha}$:
\be
\scD_\alpha \scS \ = \ \lp \chi_\alpha , \, \ldots \rp \ , \qquad \ti \scD^{\dot\alpha} \scS \ =  \ \lp \ti\chi^{\dot \alpha} , \, \ldots \rp \ .
\ee
It is straightforward to see that the $M$ component of $\scD_\alpha \scS$ and the $\ti M$ component of $\ti \scD^{\dot\alpha} \scS$ vanish, so $\scD_\alpha \scS$ and $\ti \scD^{\dot\alpha} \scS$ are complex linear multiplets.

As usual chiral multiplets satisfy $\ti\scD^{\dot\alpha} \scS = 0$, while anti-chiral multiplet satisfy $\scD_\alpha \scS=0$. 
The anti-chiral and chiral projectors are then defined as the squares of each of the spinor derivatives:
\begin{align}
\scD^\alpha \scD_\alpha \scS \ &=\ \lp 2 i  M , \, 0 , \,  \ldots \rp \; , \qquad 
\ti\scD_{\dot\alpha} \ti\scD^{\dot\alpha} \scS \ =  \ \lp -2 i  \ti M , \ . \ , \,0 , \,  \ldots \rp \ .
\end{align}
It follows from these definitions that if $\scS$ is a complex linear multiplet with $\ti M=0$, then $\ti \scD^{\dot\alpha} \scS$ is a chiral multiplet.
Similarly for a ``real'' linear multiplet $L$, $\ti \scD^{\dot\alpha} L $ is a chiral multiplet and $\scD_{\alpha} L$ is an anti-chiral multiplet.

\subsection{Curvature multiplets}\label{sec:curvature_multiplets}

From the fields in the gravity multiplet one can define a set of curvature multiplets~\cite{Cecotti:1987qe,Ferrara:1988qxa}, which represent the building blocks for the construction of (higher-derivative) supergravity invariants. 
These can also be obtained by solving the Bianchi identities in $U(1)$ superspace~\cite{Muller:1985vga}.

Using gauge symmetries present in the super-Poincar\'e algebra, we can start constructing two gauge vector multiplets. The first, that we call the R-symmetry multiplet $\mathcal V$, is associated with the Abelian gauge symmetry of $A_\mu$. Its components are \cite{Sohnius:1982fw}:
\be
\mathcal V \ =\  \lp  A_\mu\,,\, i \,\sigma^{ab}\psi_{ab}\,,\, -i \,\ti \sigma^{ab}\ti \psi_{ab}\,,\, \frac 14 \big(\hat R+ 6 \,\hat V^2\big)  \rp\,,
\ee
where $\hat R$ is the super-covariantisation of the Ricci scalar $R$ computed from the Levi-Civita connection, and $\hat V^2 = \hat V^{\mu}\hat V_{\mu}$.
 The second, called the spin connection multiplet, is associated with the non-Abelian local Lorentz invariance, whose gauge field is the spin connection.
It reads:
\be
\Omega_{ab} \ = \ \left(\omega^-_{\mu ab} \,,\, - 2  \psi_{ab}\,, \, - 2  \ti\psi_{ab}\,, \, 2  \hat F^+_{ab}  \right)\, .
\ee

For each of these gauge vector multiplets, we can build the associated chiral and anti-chiral field strength multiplets. The field strength multiplets of $\mathcal V$, denoted by $T_\alpha = \Lambda_{\alpha}(\mathcal V)\,$, $\ti T^{\dot \alpha} =  \ti\Lambda^{\dot\alpha}(\mathcal V)$, may be called the chiral and anti-chiral Ricci scalar multiplet (since they contain the Ricci scalar), while the field strength multiplets of $\Omega_{ab}$, denoted by $T_{ab\,\alpha}= \Lambda_\alpha(\Omega_{ab})\,$, $\ti T^{\dot \alpha}_{ab} =  \ti\Lambda^{\dot\alpha}(\Omega_{ab}) \,$, may be called the Riemann multiplets (as they contain the Riemann tensor). The expression of the Ricci scalar multiplets is
\bea
T_\alpha\!\!&=&\! \!\left(i \big(\sigma^{ab} \psi_{ab}\big)_\alpha \,,\,- \frac{1}{\sqrt 2}\hat F_{ab} \,\sigma^{ab}{}_{\alpha\beta} + \frac{i}{4\sqrt 2}\big(\hat R+6\hat V^2\big)\varepsilon_{\alpha\beta} \,,\, -\big(\sigma^{ab}\sigma^c\hat D^-_c \ti \psi_{ab}\big)_\alpha \right) \, ,\nn \\ [2mm]
\ti T^{\dot \alpha}\!\!&=&\!\! \left(-i\big(\ti\sigma^{ab} \ti\psi_{ab}\big)^{\dot\alpha} \,,\, - \frac{1}{\sqrt 2}\hat F_{ab}\, \ti\sigma^{ab\,\dot\alpha\dot\beta} -\frac{i}{4\sqrt 2}\big(\hat R+6 \hat V^2\big)\varepsilon^{\dot\alpha\dot\beta} \,,\,\big(\ti\sigma^{ab}\ti \sigma^c\hat D^-_c \psi_{ab}\big)^{\dot\alpha} \right) \,,\qquad
\eea
while the Riemann multiplets read
\bea
T_{ab\,\alpha} \!\!&=& \!\!\left( - 2 \psi_{ab\,\alpha}\,,\, - \frac{1}{\sqrt 2} \hat R^+_{abcd}\sigma^{cd}{}_{\alpha\beta} + \sqrt 2 i \hat F_{ab}^+\, \varepsilon_{\alpha\beta}  \,,\, -2i\big(\sigma^c\hat D^-_c \ti \psi_{ab}\big)_\alpha \right) \,, \nn \\ [2mm]
\ti T_{ab}^{\,\dot\alpha} \!\! &=& \!\! \left( - 2 \ti\psi_{ab}^{\dot\alpha} \,,\, - \frac{1}{\sqrt 2} \hat R^+_{abcd}\,\ti \sigma^{cd\,\dot\alpha\dot\beta} -\sqrt 2 i\hat F_{ab}^+\, \varepsilon^{\dot\alpha\dot\beta}\,,\, -2i\big(\ti \sigma^c\hat D^-_c \psi_{ab}\big)^{\dot\alpha} \right)\, .\qquad
\eea

From the gravity multiplet one can also construct a real linear multiplet with a frame index, $E_a$, having $\hat V_a$ as bottom component. This is called the Einstein multiplet as its top component contains the Einstein tensor. Its full expression is
\be\label{EinstMult}
E_a \ = \ \lp \hat V_a\,,\, - \frac i2 \epsilon_{abcd}\, \sigma^b \ti\psi^{cd} \,, \, \frac i2 \epsilon_{abcd} \,\ti\sigma^b \psi^{cd} \, , \,  \half  \hat E^-_{ab} - \frac i2 \epsilon_{abcd}\hat F^{+}{}^{cd}  \rp \,,
\ee
where 
\be
\hat E^-_{ab} \ = \   \hat R_{a b}^-  - \half g_{ab} \hat R^- \,
\ee
is the supercovariantized Einstein tensor of the connection $\omega^-$. 
This real linear multiplet can be embedded into a general multiplet, and later we will need its $D$ component at vanishing gravitino, $\psi_\mu = \ti\psi_\mu =0$. One can check that this is given by
\be
D[E_a] \ =\ - \nabla^2 V_a  + E^-_{ab} V^b + i\, \epsilon_{abcd} V^b \lp \nabla^c V^d - F^+{}^{cd} \rp \ .
\ee
Throughout this paper $\nabla$ denotes the standard Levi-Civita connection.

The Riemann multiplet admits a decomposition into irreducible multiplets, realising at the multiplet level the Ricci decomposition of the Riemann tensor into the Weyl tensor, the Einstein tensor and the Ricci scalar. This is obtained as follows.
As a first thing, one can check that the sigma-trace part of the Riemann multiplet is the Ricci scalar multiplet:
\be
T \ = \ -\frac{i}{2}\,\sigma^{ab}\,T_{ab}\ ,\qquad \ti T \ = \ \frac{i}{2}\,\ti \sigma^{ab}\,\ti T_{ab}\ .
\ee
Then one can introduce the projectors 
\be
\Sigma_{abcd} \ = \ -\frac{1}{6}\left(3 \, \sigma_{cd}\sigma_{ab} + \sigma_{ab}\sigma_{cd}\right)\ , \qquad
\ti\Sigma_{abcd} = -\frac{1}{6}\left(3 \, \ti\sigma_{cd}\ti\sigma_{ab} + \ti\sigma_{ab}\ti\sigma_{cd}\right)\ ,
\ee
satisfying $\ti \sigma^a \Sigma_{abcd} =0$, $\sigma^a \ti \Sigma_{abcd}=0\,$. These define new multiplets
\bea
W_{ab} &=&  \Sigma_{abcd}T^{cd} \nn \\ [2mm]
&=& \frac{1}{2}\left(T_{ab} + *T_{ab} + \frac{4i}{3} \sigma_{ab} T \right)\, ,
\, \nn
\eea
\bea
\ti W_{ab}  &=& \ti\Sigma_{abcd}\ti T^{cd} \nn \\ [2mm]
&=& \frac{1}{2}\left(\ti T_{ab} -*\ti T_{ab} - \frac{4i}{3} \ti \sigma_{ab} \ti T \right)\,,
\, 
\eea
satisfying the (anti-)self-duality conditions $*W_{ab} = W_{ab}$ and $*\ti W_{ab} = -\ti W_{ab}$, as well as $\ti \sigma^a W_{ab} = \sigma^a \ti W_{ab} = 0\,$.
These are the chiral and anti-chiral Weyl multiplets, and read
\bea
W_{ab\,\alpha} \!\!&=&\!\! \bigg(-2\Sigma_{abcd}\psi^{cd}{}_\alpha \,, -\frac{1}{\sqrt 2} \Big(\hat C_{abcd} + \frac{4i}{3}\hat \Fcs_{c[a}g_{b]d}\Big)\sigma^{cd}{}_{\alpha\beta} +\frac{\sqrt 2i}{3}\big(\hat  \Fcs_{ab} + *\hat \Fcs_{ab}\big)\varepsilon_{\alpha\beta}\,  \nn \\ [1mm] 
&& \hspace{88mm}-2i\big(\Sigma_{abcd}\sigma^e\hat D^-_e \ti \psi^{cd}\big)_\alpha  \bigg) \, ,\qquad \; \nn
\eea
\bea
\ti W_{ab}^{\dot\alpha} \!\!&=&\!\! \bigg(-2\ti\Sigma_{abcd}\ti\psi^{cd\,\dot\alpha} \,, -\frac{1}{\sqrt 2} \Big(\hat C_{abcd} - \frac{4i}{3}\hat \Fcs_{c[a}g_{b]d}\Big)\ti\sigma^{cd\,\dot\alpha\dot\beta} -\frac{\sqrt 2i}{3}\big(\hat  \Fcs_{ab} - *\hat \Fcs_{ab}\big)\varepsilon^{\dot\alpha\dot\beta} \, \nn \\ [1mm]
&& \hspace{88mm} -2i\big(\ti \Sigma_{abcd}\ti \sigma^e\hat D^-_e \psi_{ab}\big)^{\dot\alpha}\bigg) \,,\qquad\;
\eea
where $C_{abcd}$ is the Weyl tensor (defined in \eqref{DefWeylTensor}), while $\Fcs = \diff A - \frac 32 \diff V$ is the field strength of the gauge field appearing in conformal supergravity, and again a hat denotes supercovariantisation by gravitino terms whose precise form we will not need.

\begin{table}
\centering
\begin{tabular}{|c|c|c|c|c|c|}
\hline
name & symbol  & type & $\,\Delta$ & $r$ & equal to \\
\hline
& & & & &  \\ [-4mm]
R-symmetry & $\scV$ & gauge vector & 0 & 0&  
\\ [2mm]
\hline
& & & & &  \\ [-4mm]
spin connection & $\Omega_{ab}$ & gauge vector & 0 & 0 &
 \\ [2mm]
\hline
& & & & &  \\ [-4mm]
Einstein & $E_a$ & real linear & 1 & 0 & \\ [2mm]
\hline
& & & & &  \\ [-4mm]
Riemann &  \begin{tabular}{c} $T_{ab\,\alpha}$ \\ [2mm] $\ti T^{\dot\alpha}_{ab}$ \end{tabular}  & \begin{tabular}{c}chiral\\ [2mm] anti-chiral \end{tabular} & 3/2 & \begin{tabular}{c} $1$\\ [2mm] $-1$ \end{tabular} & \begin{tabular}{c}$ \Lambda_\alpha(\Omega_{ab})$ \\ [2mm] $\ti \Lambda^{\dot\alpha}(\Omega_{ab})$ \end{tabular} \\ 
& & & & &  \\ [-4mm]
\hline
& & & & &  \\ [-4mm]
Ricci scalar & \begin{tabular}{c} $T_{\alpha}$\\ [2mm]$\ti T^{\dot\alpha}$\end{tabular}  & \begin{tabular}{c}chiral\\ [2mm] anti-chiral \end{tabular} & 3/2 & \begin{tabular}{c} $1$\\ [2mm] $-1$ \end{tabular} & \begin{tabular}{c}$ -\tfrac i2 \lp \sigma^{ab} T_{ab} \rp_{\alpha},\quad \Lambda_\alpha(\scV)$ \\ [2mm]  $\;\;\,\tfrac i2\big( \ti\sigma^{ab} \ti T_{ab} \big)^{\dot\alpha} \;,\quad \ti\Lambda^{\dot\alpha}(\scV)$ \end{tabular}\\
& & & & &  \\ [-4mm]
\hline
&  & & & &  \\ [-4mm]
Weyl & \begin{tabular}{c} $W_{ab\, \alpha}$ \\ [2mm] $\ti W^{\dot\alpha}_{ab}$\end{tabular}  & \begin{tabular}{c}chiral\\ [2mm] anti-chiral \end{tabular} & 3/2 & \begin{tabular}{c} $1$\\ [2mm] $-1$ \end{tabular} & \begin{tabular}{c}$ \lp \Sigma_{abcd} T^{cd} \rp_{\alpha}$ \\ [2mm]  $\big( \ti\Sigma_{abcd} \ti T^{cd} \big)^{\dot\alpha}$ \end{tabular}  \\
 &  & & & &  \\ [-4mm]
   \hline
   \hline
\multirow{2}{*}{spinor derivative} & \multirow{2}{*}{ \begin{tabular}{c} $\scD_{\alpha}$ \\ [1mm] $\ti \scD^{\dot\alpha}$ \end{tabular}}  & \multirow{2}{*}{ - } & \multirow{2}{*}{1/2} & \multirow{2}{*}{\begin{tabular}{c} $-1$ \\ [1mm] $1$ \end{tabular}} &
   \\
 &  & & & &  \\ [2mm]
\hline
\end{tabular}
\caption{Curvature multiplets of new minimal supergravity, with their mass dimension $\Delta$ and R-charge $r$.}
\label{Multiplets}
\end{table}

Finally, recall that since $E_a$ is real linear, the multiplet $\ti \scD E_a = (\frac{i}{2} \epsilon_{abcd} \ti\sigma^b \psi^{cd} \, , \ldots )$ is chiral, while $\scD E_a = (-\frac{i}{2} \epsilon_{abcd} \sigma^b \ti\psi^{cd} \, , \ldots )$ is anti-chiral. In this way, we arrive at the following decomposition of $T_{ab}$ into its irreducible pieces:
\be\label{decompositionTab}
T_{ab} \ = \ W_{ab} + \frac{4i}{3} \sigma_{ab} T + i \epsilon_{abcd}\, \sigma^c \ti \scD E^d \ ,
\ee
and similarly for $\ti T_{ab}\,$. We also have the relations
\be\label{relTDE}
T \ =\ \frac 12 \sigma^a \ti \scD E_a\ , \qquad \ti T \ =\ -\frac 12 \ti\sigma^a  \scD E_a\ 
\ee 
and
\be\label{Rel_DT_tiDtiT}
\scD \,T \ = \ \ti \scD\, \ti T\ .
\ee

In Table \ref{Multiplets} we provide a summary of the different curvature multiplets introduced in this section, with some of their inter-relations. For later convenience, we also list their mass dimension $\Delta$ and R-charge $r$, which are defined as the mass dimension and R-charge of the respective bottom components. It is also worth noticing that the spinor derivative $\scD \scS$ of a multiplet $\scS$ with mass dimension $\Delta$ and R-charge $r$ has mass dimension $\Delta + 1/2$ and R-charge $r -1$, while $\ti \scD \scS$ has mass dimension $\Delta + 1/2$ and R-charge $r +1$.

\subsection{Supersymmetric actions}

We now discuss the different possibilities for obtaining supersymmetric actions. These can be constructed as the superspace integral of a given multiplet.\footnote{Although we do not really use the superspace formalism in this paper, it appears useful to give the action formulae in superspace before specifying their components.}  Since eventually we are interested in bosonic backgrounds, we will explicitly provide only their bosonic parts. The complete expressions (in Lorentzian signature) can be found in \cite{Sohnius:1982fw,Ferrara:1988qxa}.

\subsubsection*{$\boldsymbol{D}$-terms}

Given a general multiplet $\scS$ of mass dimension $\Delta$, R-charge $r=0 \,$ and whose bottom component is a scalar, a supersymmetric action is defined by the superspace integral
\be
S_D \ = \ \Lambda^{2-\Delta} \int \diff^4 x \, \diff^2 \theta\, \diff^2 \ti\theta\, E \, \scS \, ,
\ee
where $ E$ is the supervielbein determinant and $\Lambda$ is a constant with the dimension of a mass.
Its bosonic part is
\be
S_{D, \, {\rm bos}} \ =\ \Lambda^{2-\Delta} \int \diff^4 x\, e\, \scL_{D, \, {\rm bos}}   \ , \label{DtermAction}
\ee
where the bosonic Lagrangian reads
\be\label{DtermL}
\scL_{D, \, {\rm bos}} \ = \ D -  2\,a_{\mu} V^{\mu} \ .
\ee

\subsubsection*{$\boldsymbol{F}$-terms and $\boldsymbol{\ti F}$-terms}

Given a chiral multiplet $\Phi$ of mass dimension $\Delta$ and R-charge $r=2\,$, a supersymmetric action is obtained from the half-superspace integral
\be
S_F \ = \ \Lambda^{3-\Delta} \int \diff^4 x \, \diff^2 \theta\,  \scE\, \Phi \, ,
\ee
where $\scE$ is the supervielbein determinant for $F$-term actions.
 The bosonic part reads
\be
S_{F, \, {\rm bos}} \ =\ \Lambda^{3-\Delta} \int \diff^4 x \,e\, \scL_{F, \, {\rm bos}}  \ ,
\ee
with
\be
\scL_{F, \, {\rm bos}}  \ = \ F \, .
\ee

Similarly, for an anti-chiral multiplet $\ti \Phi$ of mass dimension $\Delta$ and R-charge $r=-2\,$, we can define the supersymmetric action
\be
S_{\ti F} \ = \ \Lambda^{3-\Delta} \int \diff^4 x \, \diff^2 \ti\theta \,\ti \scE\,\ti\Phi \ .
\ee
Its bosonic part reads
\be
S_{\ti F, \, {\rm bos}} \ =\ \Lambda^{3-\Delta} \int \diff^4 x\,e\, \scL_{\ti F, \, {\rm bos}} \ ,
\ee
with
\be
\scL_{\ti F, \, {\rm bos}}  \ = \  \ti F \, .
\ee

\medskip

Given the field strength multiplets $\Lambda_\alpha$ and $\ti\Lambda^{\dot\alpha}$ of a gauge vector multiplet,  the $F$-term Lagrangian $\scL_F\big[{\rm Tr}(\Lambda^2)\big]$ yields the self-dual Yang--Mills Lagrangian, while $\scL_{\ti F}\big[{\rm Tr}(\ti\Lambda^2)\big]$ gives the anti-self-dual Yang--Mills Lagrangian; see appendix~\ref{FieldStrength}. From the formulae in appendix~\ref{FieldStrength} one can see that $\scL_F\big[{\rm Tr}(\Lambda^2)\big] - \scL_{\ti F}\big[{\rm Tr}(\ti\Lambda^2)\big]$ is locally a total divergence term.

\medskip

There is a relation between the $D$-term Lagrangian of a general multiplet $\scS = (C, \dots )$ with $r=0$ and the $F$- and $\ti F$-term Lagrangians of its chiral and anti-chiral projections $\ti \scD^2 \scS = (2 \ti M , \ldots)$ and $\scD^2 \scS = (2M, \ldots )$. For bosonic $C$, this reads
\bea
\scL_{F}\big[ \ti \scD^2 \scS \big] &=& \,\, 2i \, \scL_{D}[\scS]
+ 2\, \nabla^\mu \big( a_\mu + i \nabla_\mu C + 2 C V_\mu \big) \, + \,{\rm fermions} \ , \label{DFrelation} \\ [2mm]
\scL_{\ti F}\big[ \scD^2 \scS \big] &=&\!\!\!  -2i \, \scL_{D}[\scS]
+ 2\, \nabla^\mu \big( a_\mu - i \nabla_\mu C + 2 C V_\mu \big)  \, + \,{\rm fermions} \ .
\eea

\section{Construction of counterterms}\label{sec:ConstrCount}

In this section we discuss local supersymmetric Lagrangians up to mass dimension four. We are first of all interested in those having precisely dimension four, as the respective actions play the role of finite counterterms in the field theory that is obtained by taking the rigid limit of supergravity. These are invariant under global scale transformations, but do not need to be Weyl invariant. 
 On the other hand, supersymmetric Lagrangians of lower dimension define dimensionful counterterms that may be used to renormalise the theory.

We first analyse the supersymmetric invariants that are made solely of the fields in the new minimal gravity multiplet. These are universal counterterms that exist for any four-dimensional supersymmetric field theory with an R-symmetry (so that it can be coupled to new minimal supergravity). Since we require gauge and diffeomorphism invariance, we can use the curvature multiplets of section~\ref{sec:curvature_multiplets} as building blocks for our supersymmetric terms.  
 
In the last part of this section, we shall consider additional matter multiplets, allowing to define further, non-universal counterterms.

\subsection{Marginal terms}\label{sec:finite_counterterms}

Marginal supersymmetric invariants are either $F$-type actions constructed from chiral multiplets with $(\Delta,r) = (3,2)$, or  $\ti F$ type actions constructed from anti-chiral multiplets with $(\Delta,r) = (3,-2)$, or $D$-type actions constructed from supermultiplets with $(\Delta,r) = (2,0)$.
  We will classify the terms constructed with the curvature multiplets introduced in section~\ref{sec:curvature_multiplets}.

Let us study the $F$-type and $\ti F$-type Lagrangians leading to marginal counterterms. For the $F$-terms, a priori there are several possibilities obtained contracting two multiplets out of $T,$ $W_{ab},$ $\ti \scD E_a$ with sigma matrices and the antisymmetric symbol $\epsilon_{abcd}\,$.\footnote{There is also $ \ti \scD^2 (\scD T)$ but this vanishes due to relation \eqref{Rel_DT_tiDtiT}.}
However, only three of these contractions are independent; this can be seen by observing that all such terms are chiral multiplets whose bottom component is a bilinear of $\psi_{ab}$, and the only independent possibilities are $\psi_{ab} \psi^{ab}$, $\epsilon^{abcd} \psi_{ab} \psi_{cd}$ and $\psi^{ab} \sigma_{bc} \psi^{c}{}_{a}$. 
We choose to work with the convenient combinations $T^2$, $ W_{ab}W^{ab}$ and $T_{ab} *\!T^{ab}$. 
 The analysis of the $\ti F$-terms is exactly the same, the conclusion being that there are only three independent terms, that can be chosen to be $\ti T^2$, $ \ti W_{ab}\ti W^{ab}$ and $ \ti T_{ab} *\!\ti T^{ab}$. 

We can now evaluate the independent terms identified above on a generic bosonic background. 
The multiplets are multiplied using appendix~\ref{app:TensorCalculus}, in particular the square of a field strength multiplet can be evaluated via eq.~\eqref{SquareFieldStrMult}. It is convenient to present the Lagrangians $\scL_F$ and $\scL_{\ti F}$ in the combinations $\scL_F + \scL_{\ti F}$ and $\scL_F - \scL_{\ti F}\,$. After setting $\psi_\mu = \ti \psi_\mu =0$, for $T^2$ and $\ti T^2$ we obtain: 
\bea\label{squareT}
\mathcal L_{F} \big[T^2 \big] + \mathcal L_{\ti F} \big[\ti T^2 \big] &=& F_{ab} F^{ab} - \frac{1}{8}\left( R + 6\, V^2\right)^2\ ,\nn \\ [2mm]
\mathcal L_{F} \big[T^2 \big] - \mathcal L_{\ti F} \big[\ti T^2 \big] &=& F_{ab}*\!F^{ab}\ .
\eea
The first provides a supersymmetrisation of $R^2$, while the second is a simple topological term.
For $T_{ab}*\!T^{ab}$ and $\ti T_{ab}*\!\ti T^{ab}$, we have
\bea
\scL_F\big[T_{ab}*\!T^{ab}\big] + \scL_{\ti F}\big[\ti T_{ab}*\!\ti T^{ab}\big] &=& R^+_{abcd}*\!R^{+\,abcd}- 8 \, F^+_{ab}*\!F^{+\,ab}\nn \\ [2mm]
&=&  \mathscr P + 4i \nabla^a \lp  R V_a - 2 R_{ab}V^b + 2 V^2 \, V_a - \epsilon_{abcd} V^b \nabla^c V^d  \rp \nn \\ [2mm]
&& - 8\, F_{ab} *\!F^{ab} + 16\, \epsilon_{abcd}\, \nabla^a V^b (F^{cd} - \nabla^c V^d) \nn \\ [2mm]
&=&  \mathscr P - 8\, F_{ab} *\!F^{ab} + {\rm tot.\,der.}\ ,
\eea
and 
\bea
\scL_F\big[T_{ab}*\!T^{ab}\big] - \scL_{\ti F}\big[\ti T_{ab}*\!\ti T^{ab}\big] &=&  R^-_{abcd} (*R^-*){}^{abcd}\nn \\ [2mm]
&=& \mathscr E + 8\,\nabla_b \left( V_a \nabla^a V^b \right) \nn \\ [2mm]
& = & \mathscr E  + {\rm tot.\,der.}\ .\label{epsilonTT}
\eea
The former contains the Pontryagin density
\be\label{def_PontryDensity}
\mathscr P \ = \  R_{abcd} *\!R^{abcd} \ ,
\ee
while the latter contains the  Euler density
\be\label{def_EulerDensity}
\mathscr E \ =\  R_{abcd} *\!R*^{\,abcd}
\ =\ R_{abcd} R{}^{cdab} - 4 R_{ab} R{}^{ba} + R^2\ 
\ee
(in these definitions, we are omitting the vielbein determinant that makes them actual densities).
In these two definitions, $R_{abcd}$ is the Riemann tensor of the Levi-Civita connection.
The contractions of Riemann tensors with torsion $R^\pm_{abcd}$ have been expressed using formulae in appendix~\ref{curvatures_omega_pm}. Here and in the following, by ``${\rm tot. \,  der.}$'' we denote the derivative of a globally defined quantity, that integrates to zero on a compact manifold with no boundary. 

Finally, the square of the Weyl multiplets reads
\bea\label{squareWab}
\scL_F\big[W_{ab}W^{ab}\big] + \scL_{\ti F}\big[\ti W_{ab}\ti W^{ab}\big]&=& C_{abcd}C^{abcd} - \frac{8}{3} \, \Fcs_{ab} \Fcs^{ab} \ , \nn \\ [2mm]
\scL_F\big[W_{ab}W^{ab}\big]  - \scL_{\ti F}\big[\ti W_{ab}\ti W^{ab}\big] &=&  C_{abcd}*\!C^{abcd} - \frac 83 \,   \Fcs_{ab}*\!\Fcs^{ab} \nn \\ [2mm]
&=& \mathscr P - \frac 83 \, \Fcs_{ab}*\!\Fcs^{ab} \nn \\ [2mm]
&=& \mathscr P - \frac 83 \, F_{ab}*\!F^{ab} + {\rm tot.\,der.} \ .
\eea
The former expression supersymmetrises the square of the Weyl tensor, while the latter is again a supersymmetrisation of the Pontryagin density.

It may be useful to notice that two further obvious terms such as $ T_{ab}T^{ab}$ and $\ti{\scD}E_a \ti{\scD}E^a$ can be re-expressed as a linear combination of the three independent terms above by using the following relations between supermultiplets, which are implied by the decomposition of the Riemann multiplet given in \eqref{decompositionTab}:
\bea\label{RelationsSquareMult}
&& T_{ab}T^{ab} \ =\  W_{ab}W^{ab} + \frac {8}{3} \,  T^2 - 4\, \ti{\scD}E_a \ti{\scD}E^a \ , \nn \\ [2mm]
&& T_{ab}T^{ab} + T_{ab} *\!T^{ab}  \ =\ 2\, W_{ab}W^{ab} - \frac 83 \, T^2 \ ,
\eea
with similar relations holding for the respective anti-chiral multiplets.
For $T_{ab}T^{ab}$ and $\ti T_{ab}\ti T^{ab}$ we get
\bea
\scL_F\big[T_{ab}T^{ab}\big] +  \scL_{\ti F}\big[\ti T_{ab}\ti T^{ab}\big]  &=& R^-_{abcd}R^{-abcd} -8 F^+_{ab}F^{+ab}\ ,\nn \\ [2mm]
\scL_F\big[T_{ab}T^{ab}\big] -  \scL_{\ti F}\big[\ti T_{ab}\ti T^{ab}\big]   &=& \frac 12 R^-_{abcd}*\!R^{-\,abcd}\ ,
\eea
where the curvatures of the connection with torsion again can be expressed using formulae in appendix~\ref{curvatures_omega_pm}.
 A term equivalent to $ \scL_{F}[\ti{\scD}E_a \ti{\scD}E^a]$ and $\scL_{\ti F}[ \scD E_a \scD E^a]$ will be presented shortly. 

Note that all Lagrangians above of the type $\mathcal L_F - \mathcal L_{\ti F}$ are topological densities. This is related to the fact that the difference between the chiral and the anti-chiral Yang--Mills Lagrangian of a gauge vector multiplet is locally a total derivative~\cite{Ferrara:1988qxa}.

\medskip

We now come to marginal gauge-invariant $D$-type actions. The possible multiplets with $(\Delta,r) = (2,0)$ are $E_a E^a$, $\scD T$ and $\ti \scD \ti T$.\footnote{There are also $\scD\sigma^a \ti \scD E_a$ and $\ti \scD \ti \sigma^a \scD E_a$, but these are redundant due to relations \eqref{relTDE}.} 
However $\scD T$ and $\ti \scD \ti T$ are complex linear multiplets, hence their $D$-term Lagrangian is a total derivative that integrates to zero on a manifold with no boundary. This can be seen from eq.~\eqref{Dterm_ComplexLinearMult}, taking $C$ to be a scalar and $r=0$. The remaining possibility, namely
$\scL_D[E_a E^a]$, is related to the $F$-term of $ \ti{\scD}E_a \ti{\scD}E^a = \frac{1}{2} \, \ti{\scD}^2(E_aE^a)$ through relation \eqref{DFrelation}. Since the derivative term in \eqref{DFrelation} integrates to zero in this case, the action obtained from $\scL_D[E^a E_a]$ is the same as the one following from $ \scL_{F}[\ti{\scD}E_a \ti{\scD}E^a]$. An explicit evaluation starting from \eqref{EinstMult} and using the formulae in appendix~\ref{curvatures_omega_pm} gives the bosonic Lagrangian
\bea
\scL_{D}[E_a E^a] \!\!&=&\!\!  - \frac 14 \left[ R^-_{ab}R^{-ab} + 8 V^a \square V_a + 4 \nabla_a V_b \nabla^a V^b  - 4 F^+_{ab}F^{+ ab} -2i\, \epsilon^{abcd}F^+_{ab}R^-_{cd} \right]\nn \\ [2mm]
\!\!&=&\!\!   -\frac{1}{4} \Big[ R_{ab}R^{ab} + 4 RV^2  - 12 \nabla_{[a}V_{b]} \nabla^a V^b + 12 V^4 - 4 F_{ab}F^{ab} + 8 F_{ab}\nabla^{a}V^{b} \Big]\nn \\ [2mm]
\!\!&&\!\! +\; \rm{tot.\,der.}\ ,\label{LDE2}
\eea
which gives a supersymmetrisation of $R_{ab}R^{ab}$ not independent of the terms given above.

\medskip
 
To summarise, we have obtained six independent supersymmetric Lagrangians satisfying the requirements; these are given in eqs.~\eqref{squareT}--\eqref{squareWab}. However, one combination is a total derivative of a globally defined quantity, and we will discard it. In Lorentzian signature, these supersymmetric terms quadratic in the curvature were already given in~\cite{Cecotti:1987qe} at the linearised level in the supergravity fields, and in \cite{Ferrara:1988qxa,deRoo:1990zm} at the non-linear level.\footnote{The Lorentzian counterparts of the $\scL_F + \scL_{\ti F}$ and $\scL_F - \scL_{\ti F}$ Lagrangians are the real and imaginary parts of the $F$-term Lagrangian. In Euclidean signature this is not necessarily the case as all bosonic fields can be complex.} We do not find additional independent terms. Integrating on a compact manifold with no boundary we obtain the five marginal actions given in eq.~\eqref{FiniteCounters}.

\subsection{Dimensionful terms}

A priori, the possibilities are Lagrangians of mass dimension two and zero. A dimension zero Lagrangian would be a supersymmetrisation of the cosmological constant, which however does not exist in new minimal supergravity; this is due to the interplay between supersymmetry and R-symmetry~\cite{Sohnius:1981tp,Sohnius:1982fw}. For the field theories obtained as a rigid limit of new minimal supergravity, this indicates the absence of quartic divergences in supersymmetric observables.

At mass dimension two, we did not find neither an $F$- or $\ti F$-term Lagrangian, nor a $D$-term Lagrangian of a globally defined multiplet that can be constructed respecting the requirements. The only term that is obtained using just the fields in the gravity multiplet is the $D$-term of the R-symmetry gauge vector multiplet $\mathcal V$, which corresponds to the Einstein--Hilbert term of new minimal supergravity~\cite{Sohnius:1981tp}. Indeed, plugging $a_{\mu} = A_{\mu}$, $D = \frac 14 (R + 6 V_{\mu} V^{\mu})$ in \eqref{DtermL}, we obtain
\be\label{LagrangianNewMin}
\scL_D [\scV] \ =\ \frac 14 \left(R + 6 V_\mu V^\mu - 8 A_\mu V^\mu \right)\,,
\ee
which is the bosonic part of the EH term. Notice that because $\nabla^\mu V_\mu=0$, the EH action \eqref{EHaction} is invariant under
 gauge transformations of the background field $A$.

\subsection{Additional background fields}

We now consider gauge vector multiplets and chiral multiplets in addition to the gravity multiplet and construct more invariant actions. Let us assume that these fields are not path integrated over in the field theory obtained by taking the rigid limit of new minimal supergravity, so that they reduce to background fields and the respective invariant actions play the role of counterterms rather than kinetic or superpotential terms of dynamical fields. 

The $D$-term of a background Abelian gauge vector multiplet $(a_\mu,\lambda,\ti\lambda,D)$ (coupling to a flavor supercurrent) defines a dimension two supersymmetric Lagrangian (since the gauge vector has mass dimension one), corresponding to a standard background FI term in curved space. Its bosonic part reads as in \eqref{DtermL}.
Moreover, from any gauge vector multiplet one can define the field strength chiral multiplet and then the associated supersymmetric Yang--Mills and topological actions. The $F$- and $\ti F$-term Lagrangians have mass dimension four, and their bosonic parts are:
\bea
\scL_F + \scL_{\ti F} &=& f_{\mu\nu}f^{\mu\nu} - 2 D^2\ , \nn \\ [2mm]
\scL_F - \scL_{\ti F} &=& f_{\mu\nu}*\!f^{\mu\nu}\ , \label{gauge_cttrm}
\eea
where $f_{\mu\nu} = 2 D_{[\mu} a_{\nu]}$ is the field strength of $a_\mu$.
It is straightforward to extend this to more background vector multiplets by pairing up the respective field strength multiplets via the multiplet tensor calculus (see appendix~\ref{FieldStrength}).

This also provides mixed gravity-gauge invariants: multiplying the Ricci scalar multiplet $T$ (recall that it is the field strength multiplet of the R-symmetry gauge vector multiplet $\scV$) with the field strength multiplet of an arbitrary Abelian gauge vector multiplet via eq.~\eqref{MixedLagrangians}, we obtain the bosonic Lagrangians:
\bea
&&\scL_{\textrm{gravity-gauge},\,F} + \scL_{\textrm{gravity-gauge},\,\ti F} \ = \ F_{\mu\nu}f^{\mu\nu} - \frac 12 \left(R + 6\, V^2\right) D\ , \nn \\ [2mm]
&&\scL_{\textrm{gravity-gauge},\,F} - \scL_{\textrm{gravity-gauge},\,\ti F} \ = \ F_{\mu\nu} *\!f^{\mu\nu}\ .\label{gravity-gauge_cttrm}
\eea

If background chiral multiplets are introduced, there are several further counterterms one can construct. 
 An example is the supersymmetrisation of the coupling $E_{\mu\nu}\partial^\mu \phi \partial^\nu \widetilde\phi$ between the Einstein tensor and a chiral multiplet presented in~\cite{Farakos:2012je}. In section~\ref{sec:PhysicalImplications} we will say more about counterterms constructed with no derivatives of chiral fields.

\section{Vanishing results on supersymmetric backgrounds}\label{sec:VanishingResults}

In this section we show that the supersymmetric counterterms largely trivialise on a bosonic background preserving supersymmetry.

\subsection{Supersymmetric backgrounds}

\subsubsection*{Review of implications of $\delta \psi_\mu = \delta \ti \psi_\mu = 0$}

A bosonic background preserves supersymmetry if besides setting all fermions to zero, we impose that their supersymmetry variations also vanish. This constrains the bosonic fields. In particular, setting $\delta \psi_\mu = \delta\ti \psi_\mu = 0$ in \eqref{SusyVarGravity}, and recalling \eqref{D+SpinorParam}, we obtain the equations
\bea
\lp \nabla_{\mu} - i A_{\mu} \rp \zeta  + i V_{\mu} \zeta + i V^{\nu} \sigma_{\mu\nu} \zeta \!& = &\! 0  \ , \label{KeqnZeta}
 \\ [2mm]
\lp \nabla_{\mu} + i A_{\mu} \rp \ti\zeta  - i V_{\mu} \ti\zeta - i V^{\nu} \ti\sigma_{\mu\nu} \ti\zeta \!& = &\! 0 \ .\label{KeqnTiZeta}
\eea
A non-zero solution (``Killing spinor'') $\zeta$ to the first equation, or $\ti \zeta$ to the second equation, determines a supercharge for the field theory defined on the background specified by the profile of the fields $e^a_\mu$, $A_\mu$, $V_\mu$ entering in the equation~\cite{Festuccia:2011ws}. Here we will assume that the vielbein, and thus the metric, take real values. It was showed in~\cite{Klare:2012gn,Dumitrescu:2012ha} that a necessary and sufficient condition for the existence of at least one non-zero solution to \eqref{KeqnZeta} or \eqref{KeqnTiZeta} is that the four-manifold is complex and the metric Hermitian. For a solution $\zeta$ to eq.~\eqref{KeqnZeta}, the complex structure is given by the spinor bilinear $J^\mu{}_\nu  =  \frac{2i}{|\zeta|^2} \, \zeta^{\dagger}\sigma^\mu{}_\nu\zeta \,$. 
One can also introduce a complex two-form as   $P_{\mu\nu}  =  \zeta \sigma_{\mu\nu}\zeta$, 
 of type $(0,2)$ with respect to $J^\mu{}_\nu\,$. It can be proven \cite{Dumitrescu:2012ha} that in an appropriate frame, the spinor solution can be expressed in terms of a  complex 
function $s$ as $\zeta_{\alpha} = \sqrt{\frac{s}{2}}${\footnotesize{$\left(\!\!\begin{array}{c} 0\\ [-2pt] 1\end{array}\!\!\right)$}}, and the supergravity auxiliary fields are determined by
\bea
V_\mu & =& -\frac{1}{2}\nabla^\nu J_{\nu\mu} + U_\mu  \ , \label{exprV_DFS}\\ [2mm]
A_\mu &=& A^{\rm c}_\mu - \frac{1}{4}(\delta_\mu^\nu - i J_\mu{}^\nu)\nabla^\rho J_{\rho\nu} + \frac{3}{2} U_\mu \  , 
\label{exprA_DFS}
\eea
where $A^{\rm c}_\mu$ is defined as
\be\label{Ac_DFS}
A^{\rm c}_\mu \ = \ \frac{1}{4} J_\mu{}^\nu\partial_\nu\log \sqrt{g} - \frac{i}{2} \partial_\mu \log s \ ,
\ee
$g$ being the determinant of the metric in complex coordinates.
The background fields contain an arbitrariness parametrised by the choice of vector field $U^{\mu}$,  which is constrained to be holomorphic, namely 
$J^{\mu}{}_{\nu}U^{\nu} = i\, U^{\mu}$, and to obey $\nabla_{\mu} U^{\mu}=0$.  
 
The analysis is completely analogous for a solution $\widetilde \zeta$ to~\eqref{KeqnTiZeta}, with the complex structure being given by $\ti J^\mu{}_\nu =  \frac{2i}{|\ti\zeta|^2} \, \ti\zeta^{\,\dagger\,}\ti\sigma^\mu{}_\nu\ti\zeta 
\,$; see \cite{Dumitrescu:2012ha} for more details.

When there exist both a non-zero solution $\zeta$ to~\eqref{KeqnZeta} and a non-zero solution $\widetilde \zeta$ to~\eqref{KeqnTiZeta}, the field theory has two supercharges of opposite R-charge. In addition to two complex structures, $J^\mu{}_\nu$, $\widetilde  J^\mu{}_\nu$, associated with $\zeta$ and $\ti\zeta$ respectively,
 in this case one can introduce the complex vector field
\be\label{KilVec}
K^{\mu} \ =\ \zeta \sigma^{\mu}\ti\zeta  \,.
\ee
This is Killing and holomorphic with respect to both complex structures. If $K^{\mu}$ commutes with its complex conjugate, 
$K^{\nu} \nabla_{\nu} \overline K{}^{\mu} - \overline K{}^{\nu} \nabla_{\nu} K^{\mu} =0$, 
then the vector field $U^\mu$ above is restricted to take the form $U^{\mu} = \kappa K^{\mu}$, where $\kappa$ is a complex function whose only constraint is to satisfy 
$K^{\mu} \p_{\mu}\kappa = 0$.

\subsubsection*{One supercharge}

We now analyse the supersymmetry conditions for a general multiplet. A similar analysis has been performed in \cite{Closset:2014uda} using twisted variables.

Let us assume that the background admits a solution $\zeta$ to eq.~\eqref{KeqnZeta}, while $\ti \zeta = 0\,$, and consider a general multiplet whose bottom component $C$ is an uncharged scalar, so that we can construct a $D$-term action. The supersymmetry conditions following from the variations \eqref{GenMultTransfo} impose $M =0$, leave $\ti M$ arbitrary, and give the equations
\bea
\ti\sigma^{\mu} \zeta \lp  i a_{\mu}  -  \partial_{\mu}C \rp &=& 0\ , \nn \\ [2mm]
i\,\sigma^{\mu\nu} \zeta \, f_{\mu\nu} &=&  \zeta D \ ,\label{EqForD}
\eea
where here $f_{\mu\nu} = 2\partial_{[\mu} a_{\nu]}\,$.
The first tells that the vector $a^\mu + i \,\partial^\mu C$ is holomorphic with respect to the complex structure $J^\mu{}_\nu\,$.
The second is equivalent to
\be\label{susy_general_mult_1}
D \ =\ \frac 12 J^{\mu\nu} f_{\mu\nu}\ , \qquad P^{\mu\nu} f_{\mu\nu} = 0\ ,
\ee 
and implies
\be\label{susy_general_mult_2}
D^2 \ =\ \frac 12 f_{\mu\nu}  f^{\mu\nu} + \frac 12 f_{\mu\nu} *\!f^{\mu\nu} \,.
\ee
Then the $D$-term action \eqref{DtermAction} evaluates to
\bea\label{evaluation_Dterm}
S_{D} & =& \int  \diff^4 x \, e \lp  \frac 12  J^{\mu\nu} f_{\mu\nu}  -  2\,a_{\mu} V^{\mu }\rp  \nn\\ [2mm]
 &=& \int  \diff \lp  J \wedge a \rp  - 2 \int  \diff^4 x \, e \, a_{\mu} U^{\mu} \  ,
\eea
where we have used the expression~\eqref{exprV_DFS} for $V$. 
 We are assuming that the manifold is compact with no boundary, so that we can integrate by parts. 
Since the contraction of two holomorphic vectors vanishes, it holds that
\be
(a_\mu + i\, \partial_\mu C)\,U^\mu \ =\ 0 \ ,
\ee
and using $\nabla_\mu U^\mu = 0$, the $D$-term action further simplifies to\footnote{This can equivalently be expressed as
$
S_D  \,= \, \int \diff^4 x\, e \, i \, \nabla^{\mu} \big( a_\mu + i \partial_\mu C + 2 C V_\mu \big) \, .
$\label{foot:DtermL}}
\be\label{evaluation_Dterm_bis}
S_{D} \ = \ \int \diff^4 x\, e \, \nabla^\mu \big( J_{\mu}{}^\nu a_\nu + 2i\, C U_\mu \big) \ .
\ee
As long as the multiplet does not have gauge redundancies, the term in parenthesis is globally defined, 
hence $S_D$ vanishes on a compact manifold with no boundary.

If instead we have an Abelian gauge vector multiplet, the conclusions are different. When working in Wess--Zumino gauge, the supersymmetry transformations are given by a combination of the transformations inherited from the general multiplet, and a gauge transformation needed to restore the gauge condition.
 The constraints imposed by supersymmetry are only $\delta_{\zeta} \lambda= \delta_{\zeta} \ti\lambda = 0$,  leading just to the second equation in \eqref{EqForD}, and therefore to \eqref{susy_general_mult_1}, \eqref{susy_general_mult_2}. Nevertheless, the condition $P^{\mu\nu} f_{\mu\nu} = P^{\mu\nu} \partial_\mu a_\nu = 0$ can be expressed as $\partial_{[i} a_{j]} =0$, where $i,j$ are holomorphic indices with respect to the complex structure $J$, and implies that {\it locally} there is a function $C(z,\bar z)$ such that $a_i = -i\,\partial_i C$. It follows that
\be
a_\mu U^\mu \ =\  a_i U^i \ =\ -i\,\partial_i C \, U^i \ =\  -i\, \partial_\mu C\, U^\mu \ =\  -i\, \nabla_\mu( C\, U^\mu)\ ,
\ee
where in the first and third equalities we used $U^{\bar \imath} = 0\,$.
 Thus the $D$-term action again takes the form \eqref{evaluation_Dterm_bis}. However, in this case $a_{\mu}$ and $C$ do not need to be globally defined, hence the action may be non-zero.

For a chiral multiplet $\Phi$, the supersymmetry conditions boil down to $F = 0\,,$ while the bottom component $\phi$ is unconstrained. If we have an anti-chiral multiplet, supersymmetry just imposes that the vector $D^\mu \ti \phi$ is holomorphic with respect to $J$, while $\ti F$ is unconstrained.
Then the $F$-term Lagrangian vanishes identically on a supersymmetric background, while the $\ti F$-term is unconstrained:\footnote{We also notice that plugging this vanishing result in \eqref{DFrelation}, we recover the expression for the $D$-term Lagrangian on a supersymmetric background given in footnote~\ref{foot:DtermL}.
}
\be
\scL_{F} \,=\, 0  \ , \qquad  \scL_{\ti F}  \;\;\, {\rm arbitrary} \ .
\ee

If we consider the chiral and anti-chiral field strength multiplets $\Lambda_\alpha$ and $\ti \Lambda^{\dot\alpha}$ of a gauge vector multiplet, we obtain that the chiral Yang--Mills Lagrangian $\scL_F\big[{\rm Tr}\,\Lambda^2\big]$ vanishes,\footnote{This is just the same as condition \eqref{susy_general_mult_2}.}
\be
\scL_F \big[{\rm Tr}(\Lambda^2)\big] \ \equiv \  \frac{1}{2} f_{\mu\nu}f^{\mu\nu} +  \frac{1}{2} f_{\mu\nu}*\!f^{\mu\nu} - D^2 \ =\ 0 \ ,
\ee
implying that the anti-chiral Yang--Mills Lagrangian $\scL_{\ti F}\big[{\rm Tr}\,\ti\Lambda^2\big]$ becomes a topological term:  
\be
\scL_{\ti F}\big[{\rm Tr}(\ti \Lambda^2)\big] \ =\ - f_{\mu\nu}*\!f^{\mu\nu}\ .
\ee
This generalises to Yang--Mills Lagrangians constructed by pairing different field strength multiplets in a straightforward way.

\medskip

If the situation is reversed, namely if we have a supercharge associated with $\ti\zeta$, while $\zeta =0$, then the $D$-term action evaluates to 
\be
S_D \ = \ \int \diff^4 x\, e\, \nabla^\mu \big( - \ti J_{\mu}{}^\nu a_\nu - 2i\, C U_\mu \big)\ ,
\ee
which again vanishes for a general multiplet, while in general is non-zero for a gauge vector multiplet. This time the $\ti F$-term Lagrangian vanishes, while the $ F$-term is unconstrained:
\be
\scL_{\ti F} \,=\, 0 \ , \qquad  \scL_{F}  \;\;\, {\rm arbitrary} \ .
\ee
In particular, the chiral and anti-chiral Yang--Mills Lagrangians evaluate to
\bea
&& \scL_{\ti F} \big[{\rm Tr}(\ti \Lambda^2)\big] \ \equiv \  \frac{1}{2} f_{\mu\nu}f^{\mu\nu} -  \frac{1}{2} f_{\mu\nu}*\!f^{\mu\nu} - D^2 \ = \ 0 \ ,\\ [2mm]
&&\scL_{F}\big[{\rm Tr}(\Lambda^2)\big] \ =\  f_{\mu\nu}*\!f^{\mu\nu}\ .
\eea

\subsubsection*{Two supercharges of opposite R-charge}

We consider now a bosonic background admitting two supercharges of opposite R-charge. It is important that the $\zeta$ and the $\ti \zeta$ transformations can be performed independently.\footnote{This is not the case in old minimal supergravity, see appendix~\ref{app:CompareOldMinimal} for a comparison.} 
Clearly this case entails more constraints compared to the one supercharge case, however these do not introduce major changes in the evaluation of the $D$-term action: again this vanishes for a general multiplet, while for an Abelian gauge vector multiplet is generically non-zero and, if the complex Killing vector $K$ introduced in \eqref{KilVec} commutes with its complex conjugate, reads
\bea
S_{D} &=& \int  \diff \lp  J \wedge a \rp  - 2 \int  \diff^4 x\, e \, \kappa\, a_{\mu} K^{\mu} \  \nn \\ [2mm]
&=&  \int \diff^4 x\, e \, \nabla^\mu \big( J_{\mu}{}^\nu a_\nu + 2i \,C \,\kappa\,K_\mu \big)\ .
\eea

For a chiral multiplet, supersymmetry requires $F = 0$ and that the vector $D^\mu \phi$ is holomorphic with respect to $\ti J$. For an anti-chiral multiplet, $\ti F = 0$ and $D^\mu \ti\phi$ is holomorphic with respect to $J$.
Since $F=\ti F = 0$, both the $F$ - and $\ti F$-term Lagrangians vanish:
\be
\scL_{\ti F} \,=\, 0 \ , \qquad  \scL_{F} \,=\, 0 \ .
\ee
In particular, both the chiral and anti-chiral Yang--Mills Lagrangians of a gauge vector multiplet vanish, implying 
\be\label{QuadraticCurvZero}
D^2 - \frac{1}{2}f_{\mu\nu}f^{\mu\nu} \ = \ 0\ ,\qquad  f_{\mu\nu} *\!f^{\mu\nu} \ =\ 0\ .
\ee

\subsection{Consequences for counterterms}\label{sec:ConseqCounter}

The analysis above shows that if we have one supercharge associated with a Killing spinor $\zeta$, then the only non-vanishing counterterms can be FI terms, i.e.\ $D$-term actions for Abelian gauge vector multiplets, taking the form \eqref{evaluation_Dterm} (or equivalently \eqref{evaluation_Dterm_bis}), or $\ti F$-terms. 
Similarly, if we have one supercharge associated with a Killing spinor $\ti \zeta$, then the only non-vanishing counterterms can be FI terms, or $F$-terms. 
If the background admits both a Killing spinor $\zeta$ and a Killing spinor $\ti \zeta$, i.e.\ two supercharges of opposite R-charge, then only FI terms can be non-zero. 

In particular, for the pure gravity terms given in section \ref{sec:finite_counterterms}, existence of one supercharge implies the relations
\bea\label{Relations_OneSupercharge}
\mathscr P - 8\, F_{ab}*\!F^{ab} + {\rm tot.\,der.} &=& \pm \,\mathscr E + {\rm tot.\,der.} \ ,\nn\\ [2mm]
C_{abcd}C^{abcd} - \frac{8}{3}\Fcs_{ab} \Fcs^{ab}  &=& \pm \mathscr P \mp \frac 83 G_{ab} *\!G^{ab}\ , \nn \\ [2mm]
F_{ab} F^{ab} - \frac{1}{8}\left( R + 6\, V^2\right)^2 &=&   \pm\, F_{ab} *\!F^{ab}\, ,
\eea
where recall that $\Fcs_{ab}*\!\Fcs^{ab} = F_{ab}*\!F^{ab} + {\rm tot.\,der.}$, and where the upper sign holds if we have a Killing spinor $\ti \zeta$, while the lower sign holds if we have a Killing spinor $\zeta$.\footnote{The first and second relations were already noted in~\cite{Cassani:2013dba}.} If we have both $\zeta$ and $\ti \zeta$, then the left and right hand sides vanish separately. 

Therefore we obtain the following relations between the marginal counterterms defined in eq.~\eqref{FiniteCounters}:
\bea\label{RelsSusyActions}
S_{\rm \mathscr P} - 8\, S_F \!&=&\! \pm\, S_E \qquad\qquad\;\;\, \Leftrightarrow \quad\;  3\,\sigma - \nu \ = \ \pm\, 2 \, \chi\ , \nn \\ [2mm]
S_{C^2} \!&=&\! \pm\, S_{\mathscr P} \mp \,\frac 83 S_F  \quad\; \Leftrightarrow \quad\; S_{C^2} \ =\ \pm\, 16 \pi^2 \Big(3 \,\sigma - \frac{1}{3}  \nu \Big) \ ,\nn \\ [2mm]
S_{R^2} \!&=&\! \pm S_F \qquad\qquad\quad \Leftrightarrow \quad\;  S_{R^2} \ =\ \pm\, 2\pi^2 \nu\ ,
\eea
where we recall that the Euler characteristic of the manifold is given by $\chi(M) = \frac{1}{32\pi^2} \int \diff^4x \, e \, \mathscr E$, the signature is $ \sigma(M) = \frac{1}{48\pi^2} \int \diff^4x \, e \, \mathscr P$, while $\nu$ is defined as $\nu = \frac{1}{2\pi^2} \int \diff^4 x \,e\, F_{\mu\nu}*\!F^{\mu\nu}$. Again, if there are two supercharges, the left and right hand sides vanish separately, so that one has $S_{C^2} = S_{R^2} = 0$, together with the topological constraints $\chi = \sigma = \nu = 0$.

We will discuss the ambiguities arising from non-zero $F$- or $\ti F$-terms in the one supercharge case in section~\ref{sec:PhysicalImplications}.

We now consider the FI terms. As we saw, these include the EH term \eqref{LagrangianNewMin}, which can be thought as an FI term for the R-symmetry gauge vector multiplet $\scV$. Let us focus on this term for definiteness, the discussion for FI terms of background non-R gauge multiplets being analogous. 
 Since the last term in (\ref{LagrangianNewMin}) can be written as $\int A \wedge H$ and is thus a Chern--Simons term involving the background R-symmetry gauge field,  the use of the  EH action  as a counterterm is insidious.
Being a Lagrangian of dimension two, its integral must be multiplied by a parameter $\Lambda^2$, where $\Lambda$ has the dimension of a mass, leading to \eqref{EHaction}. In the perspective of constructing 
 a counterterm, if $\Lambda$ is a coupling constant in the Lagrangian, it  may be thought of as the VEV of a background multiplet \cite{Seiberg:1993vc}.
 In particular, if  $\Lambda$ becomes a space-dependent field then the EH term is not invariant under R-symmetry gauge transformations, and  we 
 conclude that it cannot be used as a counterterm. For example,  this applies to the discussion of ambiguities in section \ref{Ambiguities_massive} below. 
 On the other hand, if $\Lambda$ is a UV cut-off scale in a regularisation scheme, it is less clear  whether it is correct to promote it to a background field. In this case it is more difficult to conclude if the EH term can be used to remove quadratic divergences or not; however, if these quadratic divergences arise, our analysis shows that this is the only term that can renormalise them. We will discuss the EH action further in section  \ref{evaluation}.

As a final remark, we observe that $\delta \psi_{ab} = 0$, $\delta \ti\psi_{ab} = 0$, arising as supersymmetry conditions for the Riemann multiplets, are equivalent to the integrability conditions of equations \eqref{KeqnZeta}, \eqref{KeqnTiZeta}. The projections by $\sigma^{ab}$ and $\Sigma^{abcd}$ are supersymmetry conditions for the Ricci scalar multiplet and for the Weyl multiplet, respectively. The latter coincides with the integrability condition of the charged conformal Killing spinor equation analysed in \cite{Cassani:2013dba}.

\subsection{Evaluation of the Einstein--Hilbert term}
\label{evaluation}

Assuming the existence of a solution $\zeta$ to eq.~\eqref{KeqnZeta}, and taking $a_\mu = A_\mu$ into~\eqref{evaluation_Dterm}, the Einstein--Hilbert action evaluates to
\bea\label{NewFormNewMinimal}
\Lambda^{-2}\, S_{\rm EH} &=&   2 \int \diff \left(J \wedge A \right) - 4 \int \diff^4 x \, e\, U^\mu A_\mu  \nn \\ [2mm]
&=&  2 \int \diff \left(J \wedge A^{\rm c} \right) - 4 \int \diff^4 x \, e\, U^\mu A^{\rm c}_\mu \ ,
\eea
where the second line is obtained by observing that $A_\mu - A^{\rm c}_\mu$ in \eqref{exprA_DFS} is a globally defined complex one-form of type $(0,1)$.

We can make this more explicit by considering for example the family of backgrounds studied in \cite{Assel:2014paa}. Backgrounds in this family have the form $M = S^1 \times M_3$, where $M_3$ has the topology of $S^3$. They preserve two supercharges of opposite R-charge, allow for a very general metric with $U(1)^3$ symmetry, and are labeled by two real parameters $b_1$, $b_2$, which determine the complex structure or, equivalently, the complex Killing vector $K$ as 
\be
K \ =\  \frac 12 \left( b_1 \frac{\partial}{\partial\varphi_1} + b_2 \frac{\partial}{\partial\varphi_2} - i \frac{\partial}{\partial\tau} \right)\ .
\ee 
Here $\tau$ is the coordinate on $S^1$, while $\varphi_1$, $\varphi_2$ are angular coordinates of a torus fibration over an interval parameterised by $\rho \in [0,1]$, which describes $S^3$. Since $K$ commutes with its complex conjugate, we have $U^\mu = \kappa K^\mu$.
 Moreover in \cite{Assel:2014paa} $A^{\rm c}_\mu$ was chosen globally defined, 
hence the first integral in \eqref{NewFormNewMinimal} evaluates to zero.  The 
gauge\footnote{Note that although \emph{a priori} one needs considering large gauge transformations, for the R-symmetry gauge field generically these do not exist. Firstly, large gauge transformations on a manifold with topology $S^1\times S^3$ must be necessarily along the $S^1$. We then distinguish two cases, depending on whether the gauge transformation is real or imaginary. Imaginary transformations yield a dependence $\mathrm{e}^{\tau}$ in the Killing spinor $\zeta$, which is not well-defined. 
Real transformations of the type $A \rightarrow A + n \, \diff \tau$, for $n \in \bZ$, yield a  $\mathrm{e}^{i n r \tau}$ dependence in a field with R-charge $r$, but in a theory with matter fields of 
generic (\emph{i.e.}\ irrational) R-charges, the periodicity of the fields does not allow for such large gauge transformations.}
of $A_\mu$ is chosen by requiring the spinor $\zeta$ to be independent of $\tau$, and that $A_\mu$ be regular at the poles of $S^3$. 
This fixes ${\rm Arg}(s) = {\rm sgn}(b_1) \varphi_1 + {\rm sgn}(b_2) \varphi_2$.  It was also required that $K^\mu \partial_\mu |s| = 0$. Recalling \eqref{Ac_DFS}, it follows that
\be
4\, K^\mu A^{\rm c}_\mu \ = \ 2\, K^\mu \partial_\mu {\rm Arg}(s)  \ = \ |b_1| + |b_2| \ ,
\ee
and therefore we can write 
\bea\label{S_R_intermediate}
S_{\rm EH} &=&  - 4\, \Lambda^2\int_{M} \diff^4 x \, e \,\kappa\, K^\mu A^{\rm c}_\mu \nn \\ [2mm]
&=&  -\left( |b_1| + |b_2| \right)\Lambda^2 \int_{M} \diff^4 x \,e \, \kappa \ \equiv \ \left( |b_1| + |b_2| \right) \kappa_M\, .
\label{Sfinal}
\eea
Note that the factor $ |b_1| + |b_2|$ has a clear geometrical interpretation, as it corresponds to the charge of the Killing spinor $\zeta$ along $K$, namely 
\bea
{\cal L}_K \zeta \ =\ \frac{1}{2} ( |b_1| + |b_2| ) \zeta~.
\eea
However, the complex parameter $\kappa_M$ is completely arbitrary. Thus we conclude that the EH action evaluates to an \emph{arbitrary number}, which is different from zero for generic choices of the arbitrary function $\kappa$.

\section{Ambiguities in rigid supersymmetry}
\label{sec:PhysicalImplications}

In this section we consider supersymmetric field theories defined on compact Riemannian four-manifolds using the rigid limit of new minimal supergravity. We discuss the consequences of our analysis of supergravity invariants on the characterisation of ambiguities in the field theory partition function. As explained in the introduction, the explicit evaluation of supersymmetric observables, via localization for instance, requires regularisation of UV divergences. Two choices of supersymmetric regularisation can differ by finite, local supersymmetric counterterms constructed from the background fields. These comprise the non-dynamical supergravity fields as well as the various couplings of the field theory which can be promoted to background matter multiplets. Hence the ambiguities in the partition function are characterised by the marginal supersymmetric actions built out of such background multiplets. Since we assume that the field theory can be regularised in a diffeomorphism and gauge invariant way, we also require the counterterms to be invariant under diffeomorphisms and gauge transformations of the background fields.  

\subsection{Supergravity fields}

  The vanishing results discussed in section \ref{sec:VanishingResults} show that the ambiguities depend on the number of supercharges, or Killing spinors, preserved by the background. 
We have seen that all marginal counterterms vanish on backgrounds admitting two supercharges of opposite R-charge. This implies that there is no ambiguity in the dependence of the partition function on background supergravity fields.

However on backgrounds admitting only a single Killing spinor, say an anti-chiral spinor $\ti\zeta$, the $\ti F$-terms vanish, but the $F$-terms may be non-zero. Let us first discuss the ambiguities arising from the fields in the gravity multiplet alone. The possible invariant actions are related by eqs.~\eqref{RelsSusyActions}, so we end up with only two independent non-vanishing counterterms, that are topological quantities. 
Taking them to be the Euler characteristic $\chi(M)$ and the signature $ \sigma(M)$ of the manifold, we get the counterterm
\be
S^{\rm ct}_{\rm grav} \ =\ c_1 \, \chi(M) + c_2 \, \sigma(M) \ ,
\ee
where $c_1, c_2$ are arbitrary complex numbers. Therefore the ambiguity resulting from $S^{\rm ct}_{\rm grav}$ is simply an overall complex number, except when $\chi(M) = \sigma(M) =0$, in which case there is no ambiguity. An example of complex manifold with $\chi(M) = \sigma(M) =0$ and preserving only one supercharge is the primary Hopf surface of the second type (see \emph{e.g.}\ \cite{GO})
discussed in~\cite{Closset:2013vra}.

In the presence of a background gauge vector multiplet, the situation is almost identical. We can consider the additional marginal counterterms following from integrating \eqref{gauge_cttrm}, as well as the mixed gauge-gravity Lagrangians \eqref{gravity-gauge_cttrm}.
In the presence of a solution $\ti \zeta$, the $\ti F$-terms vanish and we are left with two non-vanishing counterterms evaluating to the topological quantities given in the second lines of \eqref{gauge_cttrm} and \eqref{gravity-gauge_cttrm}. So in this case we have the extra counterterms, involving the background vector field strength $f_{\mu\nu}$:
\be\label{gauge_counterterms}
S^{\rm ct}_{\rm bkgd \, vec} \ =\ c_3 \, \int \diff^4 x \,e\, f_{\mu\nu} *\!f^{\mu\nu} + c_4 \int \diff^4 x \,e\, f_{\mu\nu} *\!F^{\mu\nu} \, ,
\ee
where $c_3, c_4$ are arbitrary complex numbers. We conclude that the ambiguity resulting from $S^{\rm ct}_{\rm grav}$ and $S^{\rm ct}_{\rm bkgd \, vec}$ is simply an overall complex number, except when all relevant topological quantities vanish, in which case there is no ambiguity.

In the presence of coupling constants, promoted to background matter multiplets, possible additional finite counterterms are given by dimension four Lagrangians built out of the gravity multiplets and the background multiplets. We consider now the potential ambiguities resulting from the presence of marginal couplings, FI terms and complex mass terms.

\subsection{Marginal couplings}

Marginal deformations of supersymmetric theories are $F$- and $\ti F$-terms constructed from (anti-)chiral superfields of mass dimension $\Delta = 3$ and R-charge $r = (-) 2$. Denoting by $\tau_I$ the complex marginal couplings associated with the pairs of (anti-)chiral superfields $\scW_I, \ti\scW_I$, we can consider the action\footnote{In this section, the superspace integrals denote actions invariant under rigid supersymmetry, so with the gravitino set to zero.}
\be
 \sum_{I} \lp \tau_I \, \int \diff^4 x \, \diff^2 \theta \ \scE \, \scW_I \ + \  \bar\tau_I \, \int \diff^4 x \, \diff^2\ti\theta \ \ti\scE  \, \ti\scW_I \rp\ .
\ee
The Yang--Mills action with complexified Yang--Mills coupling $\tau_{\rm YM} = \frac{4\pi i}{g^2_{\rm YM}} + \frac{\theta}{2\pi}$ is one example of such marginal actions. 
The marginal couplings $\tau_I$ and their complex conjugate $\bar \tau_I$ can be considered independent\footnote{In Euclidean supersymmetric theories, the chiral term $\scW_I$ and anti-chiral term $\ti\scW_I$ do not need to be related a priori. In the perspective of the analytical continuation from the Lorentzian case with real action, they come in ``conjugate'' pairs $\scW_I, \ti\scW_I$ with complex conjugate couplings. } 
 and can be promoted to background (anti-)chiral superfields of vanishing R-charge and mass dimension, since constant values of the complex scalar are supersymmetric backgrounds for an (anti-)chiral multiplet with $r=0$. Explicitly, we promote the $\tau_I$ to background chiral multiplets $\Phi_I$ and the $\bar\tau_I$ to background anti-chiral multiplets~$\ti\Phi_I$.

To address the question of the ambiguities of the partition function, we consider the possible supersymmetric counterterms that can be constructed out of $\Phi_I, \ti\Phi_I$ and the curvature multiplets discussed in the previous sections. These counterterms are $F$- and $\ti F$-term actions, evaluated on the background: 
\begin{align}
 S^{\rm ct}_{\rm marg} \ &= \ \int \diff^4 x \, \diff^2 \theta \ \scE \, g(\Phi_I) \Phi_{\rm grav} \ = \ g(\tau_I) \int \diff^4 x \, e\, \scL_F[ \Phi_{\rm grav}] \ ,  \nn\\ 
 S^{\rm ct}_{\ti{\rm marg}} \ &= \ \int \diff^4 x \, \diff^2\ti\theta \ \ti\scE  \, \ti g(\ti\Phi_I) \ti \Phi_{\rm grav} 
 \ = \ \ti g(\bar\tau_I) \int \diff^4 x \, e \,  \scL_{\ti F}[\ti \Phi_{\rm grav}] \ ,
 \label{TauCounterTerms}
\end{align}
where $g$ is an arbitrary holomorphic function of the $\Phi_I$, $\ti g$ is an arbitrary holomorphic function of the $\ti \Phi_J$ and $\Phi_{\rm grav}$, $\ti \Phi_{\rm grav}$ denote the chiral and anti-chiral combinations of curvature multiplets, with mass dimension $\Delta = 3$ and R-charges $r=2$ and $r=-2$, respectively. The analysis is just the same as when we only considered pure gravity counterterms, except that now we stick to them (anti-)holomorphic functions of the marginal couplings.

In the case of two supercharges of opposite R-charge, we have shown that the $F$- and $\ti F$ actions vanish identically on a supersymmetric background, hence all possible counterterms vanish and the dependence on $\tau_I$ is unambiguous. 
In \cite{Closset:2014uda} it was explained that in the case of two supercharges the $F$- and $\ti F$-term actions are always $Q$-exact, meaning that they can be expressed as the supersymmetry variation of other quantities. This implies that the partition function does not depend on $\tau_I,$ $\bar \tau_I\,$, up to anomalies and counterterms. Here we have shown that this independence is not spoiled by possible counterterms.

In the case of one supercharge, say $\ti\zeta$, only the $\ti F$ actions vanish identically. The dependence on the $\bar\tau_I$ parameters is then unambiguous. It was shown in \cite{Closset:2014uda} that the $\ti F$ actions are $Q$-exact, implying that the partition function depends holomorphically on the $\tau_I$ couplings, again up to anomalies and counterterms. We have demonstrated that counterterms do not spoil this result.
On the other hand, the holomorphic dependence on the $\tau_I$ is subject to the ambiguities due to the $F$-type counterterms in \eqref{TauCounterTerms}. These are constructed with the chiral curvature multiplets of section \ref{sec:ConstrCount}, whose $F$-term actions evaluate to only two independent topological quantities, that can be chosen to be  $\chi(M)$ and $\sigma(M)$.
We conclude that the partition function has an ambiguous  dependence on the $\tau_I$ parameters characterised by the counterterms
\begin{align}
 S^{\rm ct}_{\rm marg} &= \  g_1(\tau_I) \, \chi(M) + g_2(\tau_I) \, \sigma(M)\ ,
\end{align}
where $g_1,g_2$ are arbitrary holomorphic functions of the $\tau_I$.\footnote{This does not mean that the dependence on the $\tau_I$ is always totally ambiguous. For instance, if the partition function also depends on geometric parameters (like complex structure moduli \cite{Closset:2013vra}), it can be an intricate function of the marginal parameters and these other parameters, so that the full dependence on the $\tau_I$ cannot be removed by the counterterms we described.}
We see that the ambiguity is removed in the specific case when $\chi = \sigma =0$. 

This extends to the case when background gauge vector multiplets are present in a straightforward way: in this case arbitrary functions of the $\tau_I$ multiply the topological terms in \eqref{gauge_counterterms}.

\subsection{Relevant couplings}
\label{Ambiguities_massive}

Supersymmetric gauge theories in four dimensions also admit deformations by relevant couplings. 
Let us show in one example that if these are controlled by parameters that can be promoted to background multiplets, then we do not find any ambiguities in the dependence of {\it e.g.}\ the partition function on such parameters.

We consider an Abelian gauge theory (the generalisation to the non-Abelian case being straightforward) and a pair of chiral multiplets $\Phi_+, \Phi_-$ with mass dimension one, opposite gauge charges and R-charges $r_+, r_-$ satisfying $r_+ +r_- =2$. Then one can build the complex mass term
\be
S_{\rm c.m.} \ =\ m \, \int \diff^4 x \, \diff^2\theta\, \scE \, \Phi_+ \Phi_-
\ + \ \bar m \, \int \diff^4 x \, \diff^2 \ti\theta \, \ti\scE \, \ti\Phi_- \ti\Phi_+   \ ,
\ee
where $m$ is the complex mass, which can be promoted to a chiral multiplet $\scM$ of vanishing R-charge and mass dimension $\Delta =1$ (similarly, $\bar m$ is promoted to an anti-chiral multiplet $\ti \scM$). 
We find that there is no $F$- or $\ti F$-type counterterm one can construct by multiplying a combination of the curvature multiplets by positive powers of $\scM, \ti\scM$. We do not allow for negative powers of $\scM, \ti\scM$, since they would be singular at $m=0$, so we do not need to consider further higher-derivative counterterms.
We also do not find any $D$-type counterterm that is invariant under local R-symmetry transformations. Generically, with a background field $\scS$ of mass dimension two and vanishing R-charge (like $\scM^2$), one might be tempted to consider the $D$-term of $\scS \scV$, where we recall that $\scV$ is the R-symmetry gauge vector multiplet, but as already observed in section \ref{sec:ConseqCounter} this is not invariant under local R-symmetry transformations, so we do not allow for it.
We conclude that there is no ambiguity in the dependence on $m$ and $\bar m$, whatever number of supercharges is preserved by the background.

A similar argument may be applied to the parameter appearing in the FI term for a dynamical Abelian gauge vector multiplet in the field theory. However, in various situations this parameter has to be quantised, so it is less clear if it can be promoted to a field. The field theory FI term is given by
\be
S_{\rm FI} \ =\ \xi \, \int \diff^4 x \, \diff^2 \theta \, \diff^2 \ti\theta \ E \, \scA 
\ = \  \xi \, \int \diff^4 x\, e \, \lp D - 2 a_\mu V^\mu \rp \ ,
\label{FIterm}
\ee  
where $\xi$ is the FI parameter, with mass dimension two, and $\scA = (a_\mu , \ldots, D)$ is a dynamical Abelian gauge multiplet.\footnote{FI terms in supersymmetric field theories and supergravity have been thoroughly discussed in \cite{Komargodski:2009pc}.}
If the gauge group is compact and the four-manifold has non-trivial homotopy group $\pi_1(M) \neq \{0 \}$, requiring invariance of the action under large gauge transformations of $a_\mu$ imposes the quantisation of $\xi$ (see \emph{e.g.} \cite{Aharony:2013dha}). In particular, in the case when the manifold has topology
 $S^1 \times S^3$, with $\tau \in [0,2\pi]$ parametrising the $S^1$, invariance under the gauge transformation $a \rightarrow a + \diff \tau $ \footnote{Here we assume that the $U(1)$ charges of the matter fields of the theory are integers.} leads to the quantisation condition 
 $\xi = k \, \xi_0$, with  $k \in \bZ$ and
 \be
  \, \pi i \, \xi_0^{-1} \ =\  \int \diff ^4 x \, e \, V^{\tau}\ .
  \label{quantcond}
 \ee
  Note that in general this condition depends on the gravity background and in the specific cases when $V$ vanishes, the FI parameter is \emph{a priori} not quantised. Moreover \eqref{quantcond} leads in general to a complex FI parameter, which is somehow non-standard, however in various cases, for instance in \cite{Assel:2014paa} for theories on $S^1 \times S^3$, it was found that $V^\tau$ can be chosen purely imaginary, implying that the FI parameter is real.
  
When it is not quantised, the FI parameter may be promoted to a field; then the argument used in the example of the complex mass implies that this does not lead to an allowed counterterm. When the FI parameter is quantised, it may be meaningful to consider a term given by $\xi_0$ multiplying the Einstein--Hilbert term, however this product is not the integral of a local density, hence it does not appear to parametrise an ambiguity.

We can conclude that there are no ambiguities in the dependence of the partition function on relevant couplings.

\section{Implications for anomalies}\label{sec:anomalies}

In this section we discuss the consequences of the relations given in section~\ref{sec:VanishingResults} for the Weyl anomaly and the chiral anomaly of the R-current and flavor currents, expanding the results of \cite{Cassani:2013dba}.

We consider an $\mathcal N=1$ superconformal field theory on a general curved four-manifold. The background gravity multiplet of new minimal supergravity couples to the R-current multiplet (this contains R-current $J_{\rm R}^\mu$ as well as the energy-momentum tensor $T_{\mu\nu}$). We also introduce a background $U(1)$ gauge vector multiplet coupling to the supercurrent of a non-R flavor symmetry (the generalisation to more than one background gauge vector multiplets coupling to different flavor supercurrents is straightforward).
In a supersymmetric theory the Weyl and chiral anomalies can be written in terms of supermultiplets. Here we will provide their bosonic expressions, and discuss the consequences of imposing supersymmetry of the background.

The Weyl and R-current anomalies read \cite{Anselmi:1997am,Anselmi:1997ys,Cassani:2013dba}:
\bea
\langle T_\mu^\mu \rangle \, = \, \frac{c}{16\pi^2} \Big( C_{\mu\nu\rho\sigma}C^{\mu\nu\rho\sigma} - \frac 83 \Fcs_{\mu\nu}\Fcs^{\mu\nu} \Big) \,- \,
\frac{a}{16\pi^2} \mathscr E + \frac{3}{32\pi^2}{\rm Tr}(\mathbf{RF}^2) \left(f_{\mu\nu}f^{\mu\nu} - 2D^2 \right) \,,\label{Weyl_anomaly}
\eea
\bea
\langle \nabla_\mu J_{\rm R}^\mu \rangle \ = \  \frac{c-a}{24\pi^2}\, \mathscr P\, + \,\frac{5a-3c}{27\pi^2} \,G_{\mu\nu} *\! G^{\mu\nu} + \frac{1}{16\pi^2} {\rm Tr}(\mathbf{RF}^2) f_{\mu\nu}*\! f^{\mu\nu}\, ,\label{R_anomaly}
\eea
where $a$ and $c$ are the central charges of the superconformal theory, expressed in terms of the `t Hooft anomalies for the R-charges $\mathbf{R}$ of the fermions as \cite{Anselmi:1997am} $a = \frac{3}{32} ( 3 \, \textrm{Tr}\, \mathbf{R}^3  - \textrm{Tr} \,\mathbf{R} )$ and $c = \frac{1}{32} ( 9 \,  \textrm{Tr}\, \mathbf{R}^3  - 5 \, \textrm{Tr} \,\mathbf{R} )$, while ${\rm Tr}(\mathbf{RF}^2)$ is a mixed 't Hooft anomaly involving the charges $\mathbf{F}$ of the fermions under the flavor symmetry. Moreover, $f_{\mu\nu}$ is the field strength of the background gauge field coupling to the flavor current.
The reason why the same coefficients $a$, $c$ and ${\rm Tr}(\mathbf{RF}^2)$ appear in the Weyl and R-symmetry anomalies is because the trace of the energy-momentum tensor and the divergence of the R-current are part of the same supertrace multiplet, and the two equations \eqref{Weyl_anomaly}, \eqref{R_anomaly} arise from a single super-anomaly equation. Generically also the flavor current acquires anomalous contributions from the background fields, so that its divergence reads~\cite{Intriligator:2003jj}:\footnote{We are requiring there is no contribution from the dynamical fields in the theory.}
\be
\langle \nabla_\mu J_{\rm f\,\!l}^\mu \rangle  =   \frac{1}{384\pi^2} {\rm Tr}(\mathbf{F}) \Big(\mathscr P - \frac{8}{3} G_{\mu\nu}* G^{\mu\nu} \Big) \, +  \frac{1}{16\pi^2}\, {\rm Tr}(\mathbf{RF}^2) F_{\mu\nu}* f^{\mu\nu}  \, +  \frac{1}{48\pi^2} {\rm Tr}(\mathbf{F}^3) f_{\mu\nu}*\! f^{\mu\nu}.
\ee

When the background preserves two supercharges of opposite R-charge, by using the relations derived in section \ref{sec:VanishingResults} we obtain that all anomalies above become at most a total derivative:
\bea
\langle T_\mu^\mu \rangle &=& a \ {\rm tot.\,der.}\ ,\nn \\ [2mm]
\langle \nabla_\mu J_{\rm R}^\mu \rangle &=&  a \ {\rm tot.\,der.} \ ,\nn \\ [2mm]
\langle \nabla_\mu J_{\rm f\,\!l}^\mu \rangle &=& 0 \ ,
\eea
so all anomalies vanish upon integration on a compact manifold with no boundary: 
\be
\int \diff^4 x \,e\, \langle T_\mu^\mu \rangle \ = \ \int \diff^4 x \,e\, \langle \nabla_\mu J_{\rm R}^\mu \rangle \ = \ \int \diff^4 x \,e\, \langle \nabla_\mu J_{\rm f\,\!l}^\mu \rangle \ = \ 0\ .
\label{Anom2KS}
\ee

When only one supercharge, say the one associated with $\ti\zeta$, is preserved, we do not obtain useful constraints on the chiral anomalies, however we observe that using the relations in section \ref{sec:VanishingResults} the Weyl anomaly can be written in terms of topological densities only. Hence its integral depends on the background fields only through the topology of the respective bundles:
\be
 \int \diff^4 x \, e\, \langle T_\mu^\mu \rangle   \ = \    3 c \,\sigma - 2a \, \chi -  \frac{c}{3}\,  \nu + \frac{3}{32\pi^2}{\rm Tr}(\mathbf{RF}^2) \int \diff^4 x\,e\, f_{\mu\nu}*\!f^{\mu\nu} \ .
 \label{Anom1KS}
 \ee

This extends the results of \cite{Cassani:2013dba} to the case when background gauge vector multiplets, coupling to flavor supercurrents, are present.

It is simple to see that the integrated anomalies \eqref{Anom2KS}, \eqref{Anom1KS} are supersymmetric observables
free of ambiguities. For instance, the only supersymmetric counterterm in \eqref{FiniteCounters} that can affect $\langle T_\mu^\mu \rangle$ is $S_{R^2}$, 
the others being topological or  Weyl invariant. This counterterm yields a contribution proportional to $\nabla^2 (R+6 V^2)$, which vanishes upon integration over the manifold.

\section{Concluding remarks}\label{sec:conclusions}

In this work we have analysed systematically local  invariants of new minimal supergravity in Euclidean signature, that may be used as counterterms for rigid $\scN=1$ supersymmetric quantum field theories defined on Riemannian four-manifolds. 

One of the main outcomes of this analysis is that when evaluated on a  background  admitting two supercharges of opposite chirality (and R-charge), all marginal 
supersymmetric counterterms vanish, implying that supersymmetric partition functions do not suffer from scheme-dependent ambiguities. These include the partition functions on Hopf surfaces computed by the authors~\cite{Assel:2014paa}, and their generalisations considered in~\cite{Nishioka:2014zpa}, as well as the partition function on $\mathbb{T}^2 \times \Sigma$, where $\Sigma$ is a Riemann surface. Partial results in the case that $\Sigma = \mathbb{C}P^1$ have been presented in \cite{Closset:2013sxa} and~\cite{Nishioka:2014zpa}.
In the case when only one supercharge is preserved, we have shown that ambiguities arise. However,  if $\chi(M)=\sigma(M)=0$ (and there are no background gauge vectors), again all marginal  counterterms vanish, leaving an unambiguous result. 
As an example, it would be interesting to  study the case of Hopf surfaces of second type  \cite{Closset:2013vra}.

We have found that the only dimension two supersymmetric invariants are FI terms (as we saw, these include the Einstein--Hilbert term), and discussed some of their features. As a result, we expect that quadratic divergences in supersymmetry-preserving regularisations are very constrained.

The  fact that finite supersymmetric counterterms vanish in the case of Hopf surfaces indicates that the supersymmetric Casimir energy defined in~\cite{Assel:2014paa} is not ambiguous and therefore is physically significant. It would be interesting to study further this Casimir energy, in particular  to investigate whether one can understand its universal nature, based on general principles, for example along the lines of \cite{Cappelli:1988vw,Herzog:2013ed,Huang:2013lhw} and~\cite{DiPietro:2014bca}.

Finally, it should be interesting to classify supersymmetric counterterms in five and six  dimensional supergravity theories.

\subsection*{Acknowledgments}

We would like to thank  L.~Di Pietro and Z.~Komargodski for constructive comments on a first draft of this paper.
We are also grateful  to C.~Closset, S.~Cremonesi, J.~Gomis, M.~Porrati and P.~West for interesting discussions.
B.A. and D.M. are supported by the ERC Starting Grant N. 304806, ``The Gauge/Gravity Duality and Geometry in String Theory''.
D.C. is supported by an European Commission Marie Curie Fellowship under the contract PIEF-GA-2013-627243. D.C. acknowledges previous support by the STFC grant ST/J002798/1 while he was a Research Associate at King's College London, where part of this work was done.
D.C. would like to thank CERN, and also the CERN-Korea Theory Collaboration funded by National Research Foundation (Korea), for hospitality and support during the workshop ``Exact Results in SUSY Gauge Theories in Various Dimensions''.

\appendix

\section{Conventions and useful identities}\label{app:conventions}

We use Greek letters $\mu,\nu,\ldots$ for curved space indices and Latin letters $a,b,\ldots$ for frame indices.
We work in Euclidean signature $(++++)$. 
The totally antisymmetric symbol $\epsilon_{abcd}$ is normalised as $\epsilon_{1234} = 1$. For any antisymmetric rank-two tensor $t_{ab}$, we define the Hodge dual $* t_{ab} = \frac 12 \epsilon_{abcd}\,t^{cd}$.

\subsection{Spinors}

We adopt a two-component spinor notation: positive chirality spinors carry an undotted index, as $\zeta_\alpha$, $\alpha =1,2$, 
while negative chirality spinors are distinguished by a tilde and carry a dotted index, as $\widetilde \zeta^{\dot\alpha}$. The Hermitian conjugate spinors are
$(\zeta^{\dagger})^{\alpha}= (\zeta_{\alpha})^{\ast} \,,$   $(\ti\zeta^{\dagger})_{\dot\alpha}= (\ti\zeta^{\dot\alpha})^{\ast}$,
and the spinor norms are given by $|\zeta|^2 \,=\, \zeta^{\dagger \, \alpha}\zeta_{\alpha}$ and $|\ti\zeta|^2 \,=\, \ti\zeta^{\,\dagger}_{\dot\alpha}\,\ti\zeta^{\dot\alpha}\,$. We will often omit the explicit spinor indices.

The Clifford algebra is generated by sigma matrices
\be
\sigma^a_{\alpha \dot\alpha} \; =\; (\vec\sigma , -i\mathbbm{1}_2) \, ,\qquad \qquad \ti\sigma^{a \, \dot\alpha \alpha}\; =\; (-\vec\sigma , -i\mathbbm{1}_2)  \ ,
\ee
where $\vec\sigma = (\sigma^1,\sigma^2,\sigma^3)$ are the Pauli matrices. These satisfy
\be
\sigma_{a} \ti\sigma_{b} + \sigma_{b} \ti\sigma_{a} \;=\; -2 \delta_{ab} \, , \qquad\quad  \ti\sigma_{a} \sigma_{b} + \ti\sigma_{b} \sigma_{a} \;=\; -2 \delta_{ab}\ .
\ee
 The matrices
\begin{align}
\sigma_{a b} \,&=\, \frac{1}{4} \lp \sigma_{a} \ti\sigma_{b} - \sigma_{b} \ti\sigma_{a}  \rp \,, \qquad 
\ti\sigma_{a b} \,=\, \frac{1}{4} \lp \ti\sigma_{a} \sigma_{b} - \ti\sigma_{b} \sigma_{a}   \rp  
\end{align}
are self-dual and anti-self-dual, respectively 
\be
\half \epsilon_{abcd} \, \sigma^{cd} \,=\, \sigma_{ab}\,,\qquad\qquad \half \epsilon_{abcd}\, \ti\sigma^{cd} \,=\, -\ti\sigma_{ab}\, .
\ee 
The sigma matrices have the following hermiticity properties
\be
(\sigma_{a})^{\dagger} \,=\, - \ti\sigma_{a} \, , \qquad (\sigma_{ab})^{\dagger} \,=\, - \sigma_{ab} \, , \qquad (\ti\sigma_{ab})^{\dagger} \,=\, - \ti\sigma_{ab}\;,
\ee
and satisfy
\bea
&& \sigma_{a} \ti\sigma_{b}\sigma_{c} \;=\; - \delta_{ab} \sigma_{c} 
+ \delta_{ac} \sigma_{b} - \delta_{bc} \sigma_{a} + \epsilon_{abcd} \sigma^d\,, \nn \\ [1mm]
&& \ti\sigma_{a} \sigma_{b} \ti\sigma_{c} \;=\; - \delta_{ab} \ti\sigma_{c} 
+ \delta_{ac} \ti\sigma_{b} - \delta_{bc} \ti\sigma_{a} - \epsilon_{abcd} \ti\sigma^d \,,\nn \\ [1mm]
&& \sigma_{ab}\sigma_{cd} \ =\ \tfrac{1}{4} \lp - \epsilon_{abcd} - 2 \delta_{ad}\sigma_{bc}+ 2 \delta_{ac}\sigma_{bd} - 2 \delta_{bc}\sigma_{ad} + 2 \delta_{bd}\sigma_{ac} - \delta_{ac}\delta_{bd} + \delta_{ad}\delta_{bc} \rp \,,\nn \\ [1mm]
&& \ti\sigma_{ab} \ti\sigma_{cd} \ =\ \tfrac{1}{4} \lp + \epsilon_{abcd} - 2 \delta_{ad} \ti\sigma_{bc}+ 2 \delta_{ac} \ti\sigma_{bd} - 2 \delta_{bc} \ti\sigma_{ad} + 2 \delta_{bd} \ti\sigma_{ac} - \delta_{ac}\delta_{bd} + \delta_{ad}\delta_{bc} \rp\,,\qquad \nn\\ [1mm]
&& \sigma_a \ti \sigma_{bc} \ = \ - \delta_{a[b}\sigma_{c]} + \tfrac 12 \epsilon_{abcd} \sigma^d \ , \qquad\quad \ti \sigma_a \sigma_{bc} \,\ = \ - \delta_{a[b}\ti \sigma_{c]} - \tfrac 12 \epsilon_{abcd} \ti \sigma^d\ .
\eea

We take the supersymmetry parameters $\zeta$, $\ti\zeta$ to be commuting spinors, with the supersymmetry variation $\delta_\zeta$, $\delta_{\ti\zeta}$ being Grassmann-odd operators. On the other hand, the dynamical spinor fields (including the gravitino) are assumed anti-commuting. Note that this yields a minus sign when $\delta_\zeta$ or $\delta_{\ti\zeta}$ passes through an anti-commuting spinor.
Undotted spinor indices are raised or lowered acting from the left with the antisymmetric symbols
$\varepsilon^{\alpha \beta}$ and $\varepsilon_{\alpha \beta}$, chosen such that $\varepsilon^{12} = -\varepsilon_{12} = 1$; for instance, $\zeta^\alpha = \varepsilon^{\alpha\beta}\zeta_\beta$ and $\zeta_{\alpha} = \varepsilon_{\alpha\beta}\zeta^\beta$. The same convention holds for dotted spinors, using $\varepsilon^{\dot \alpha\dot \beta}$ and $\varepsilon_{\dot \alpha\dot \beta}\,$. In a spinor bilinear, the indices are contracted as $\zeta \chi = \zeta^\alpha \chi_\alpha$ and $\ti\zeta \,\ti \chi = \ti\zeta_{\dot\alpha} \,\ti\chi^{\dot\alpha}$.
One has the following relations for commuting spinors
\bea
&& \zeta \chi \,=\, -\chi \zeta  \ , \qquad\qquad\qquad\quad \;\ti\zeta \,\ti\chi \,=\, - \ti\chi \,\ti\zeta \ ,  \nn\\ [2mm]
&& \zeta \sigma_{a} \ti\chi \,=\, \ti\chi \,\ti\sigma_{a} \zeta\ , \qquad\qquad\qquad \zeta \sigma_{ab}\chi \,=\,  \chi \sigma_{ab} \zeta \ ,\nn\\ [2mm]
&& (\sigma_{a} \ti\zeta\,)\, \chi \,=\, - \ti\zeta\, \ti\sigma_{a} \chi \ ,\qquad\quad\quad (\sigma_{ab} \zeta)\, \chi \,=\,\, - \zeta \sigma_{ab} \chi\ ,
\eea
as well as the Fierz identities
\bea\label{fierce}
(\chi_1\chi_2)(\ti\chi_3 \ti\chi_4) &=& - \tfrac 12 (\chi_1 \sigma_{a} \ti\chi_4)(\chi_2 \sigma^{a} \ti\chi_3)\ ,\nn \\ [2mm]
(\chi_1\chi_2)(\chi_3\chi_4) &=& -(\chi_1\chi_3)(\chi_4\chi_2) - (\chi_1\chi_4)(\chi_2\chi_3)\ .
\eea
The second identity also holds for dotted spinors. We will also use
\bea\label{MoreFierz}
(\chi_1  \sigma_a \ti \chi_2) \sigma^{ab}{}_\alpha{}^\beta &=& -\chi_{1\alpha} (\ti\chi_2 \ti\sigma^b)^\beta + \frac 12 (\chi_1  \sigma^b \ti\chi_2)\,\delta_\alpha{}^\beta\ ,\nn\\ [2mm]
(\chi_1 \sigma_a \ti \chi_2) \ti\sigma^{ab}{}^{\,\dot \alpha}{}_{\dot \beta} &=& (\ti\sigma^b\chi_1)^{\dot\alpha}\,\ti \chi_{2\,\dot\beta} - \frac 12 ( \chi_1  \sigma^b \ti\chi_2)\,\delta^{\dot \alpha}{}_{\dot \beta}\ .
\eea
In all these Fierz identities, one has to include an extra minus sign whenever the relation involves swapping two  anti-commuting spinors. 

\subsection{Spin connection and curvature}

The standard spinor covariant derivative is given by
\be
\nabla_\mu \zeta \ = \  \partial_\mu \zeta - \frac{1}{2}\omega_{\mu ab}\sigma^{ab}\zeta  \ ,\qquad\quad \nabla_\mu\ti\zeta \ = \ \partial_\mu \ti\zeta - \frac{1}{2}\omega_{\mu ab}\ti\sigma^{ab}\ti\zeta \ ,
\ee
where $\omega_{\mu ab}$ is the Levi-Civita spin connection. This has no torsion, $\partial_{[\mu} e_{\nu]}{}^a + \omega_{[\mu}{}^{ab}  e_{\nu]b} =0\,$, and is obtained from the vielbein $e^a_\mu$ and its inverse $e^{\mu}_a$ as 
\be\label{def_spin_conn}
\omega_\mu{}^{ab}(e) \ =\  2\,e^{\nu[a}\partial_{[\mu}e_{\nu]}{}^{b]} - e^{\nu[a} e^{b]\rho} e_{\mu c}\, \partial_\nu e^c_\rho\ .
\ee 

More generally, a spin connection with torsion satisfies $\partial_{[\mu} e_{\nu]}{}^a + \omega_{[\mu}{}^{ab}  e_{\nu]b}  = \frac 12 T_{\mu\nu}{}^a $, where $T_{\mu\nu}{}^a = - T_{\nu\mu}{}^a$ is the torsion tensor. The solution to this equation is $\omega_\mu{}^{ab} = \omega(e)_\mu{}^{ab}+ K_\mu{}^{ab}$, where $K_\mu{}^{ab}$ is the contortion tensor, related to the torsion as $K_{\mu\nu\rho} = \frac 12 (-T_{\mu\nu\rho} + T_{\mu\rho\nu } + T_{\nu\rho\mu})\, .$

From the spin connection (possibly with torsion) we can construct the Riemann tensor,
\be\label{defRiemann}
R_{\mu\nu ab} \ = \ \partial_\mu \omega_{\nu ab} -\partial_\nu \omega_{\mu ab} + \omega_{\mu a}{}^c \omega_{\nu cb} - \omega_{\nu a}{}^c \omega_{\mu cb} \ .
\ee
The Ricci tensor and Ricci scalar are defined as $R_{\mu\nu} = R^\rho{}_{\mu\rho\nu}$ and $R = g^{\mu\nu}R_{\mu\nu}$, respectively. One should note that if a connection has torsion, then $R_{\mu\nu\rho\sigma} \neq R_{\rho\sigma\mu\nu}$, $R_{[\mu\nu\rho]\sigma} \neq 0\,$, and $R_{\mu\nu}\neq R_{\nu\mu}$.
On the Riemann tensor, $*R_{abcd}$ will mean that the pair of indices being dualised is the first one, while $*R*_{abcd}$ indicates that the Hodge dual is taken both on the first and the second pair of indices. The same applies to the Weyl tensor $C_{\mu\nu\rho\sigma}$, which is defined by
\bea\label{DefWeylTensor}
C_{\mu\nu\rho\sigma} &=& \frac 12 \left( R_{\mu\nu\rho\sigma} + *R*_{\rho\sigma\mu\nu} \right) - \frac 16 R \, g_{\mu[\rho} g_{\sigma]\nu} \nn \\ [1mm]
&=& R_{\mu\nu\rho\sigma} - R_{\mu [\rho}g_{\sigma]\nu} + R_{\nu [\rho}g_{\sigma]\mu} + \frac 13 R \, g_{\mu[\rho} g_{\sigma]\nu} \ .
\eea

For the Levi-Civita connection, it holds that
\be
[\nabla_{\mu},\nabla_{\nu}] \zeta \, =\, -\half R_{\mu\nu ab} \sigma^{ab} \zeta\ , \qquad [\nabla_{\mu},\nabla_{\nu}] \ti \zeta \, =\, -\half R_{\mu\nu ab} \ti\sigma^{ab} \ti\zeta\ .
\ee

\subsection{Curvature of $\omega^\pm$}\label{curvatures_omega_pm}

In new minimal supergravity, the connections with torsion $\omega^\pm_\mu{}^{ab} = \hat \omega_\mu{}^{ab} \pm \hat H_\mu{}^{ab}$ are used. In the following, we consider their bosonic part by setting the gravitino terms to zero, which leaves us with $\omega^\pm_\mu{}^{ab} = \omega_\mu{}^{ab}(e) \pm H_\mu{}^{ab}$, and  express their curvature tensor in terms of the one associated with the Levi-Civita connection $\omega_\mu{}^{ab}(e)$ given in \eqref{def_spin_conn}.

We find:
\bea\label{curvature_Htorsion}
R^\pm_{\mu\nu\rho\sigma} \!\!&=&\!\! R_{\mu\nu\rho\sigma} \mp 2i  \nabla_{[\mu} V^\kappa \epsilon_{\nu] \rho\sigma\kappa} 
-\, 2 \left( V_\rho V_{[\mu}\, g_{\nu]\sigma} - V_\sigma V_{[\mu}\, g_{\nu]\rho}  -V^2  g_{\rho[\mu} g_{\nu]\sigma} \right),\nn \\ [2mm]
R^\pm_{\mu\nu} \!\!&=&\!\! R_{\mu\nu} \pm i\, \epsilon_{\mu\nu\rho\sigma} \nabla^\rho V^\sigma - 2\left( V_\mu V_\nu - g_{\mu\nu} V^2\right)\,,\nn \\ [2mm]
R^\pm \!\!&=&\!\! R + 6\, V^2\ ,
\eea
where $\nabla$ is the Levi-Civita connection, and $R_{\mu\nu\rho\sigma}$ its Riemann tensor, and we recall that $V_\mu$ is related to $H_{\mu\nu\rho}$ as in \eqref{HdualV}.
The square of these tensors evaluates to
\bea
R^\pm_{\mu\nu\rho\sigma}R^\pm{}^{\mu\nu\rho\sigma} \!\!&=&\!\! R_{\mu\nu\rho\sigma}R^{\mu\nu\rho\sigma} - 4 \left( 2R_{\mu\nu}V^\mu V^\nu - R V^2 + 2 \nabla_\mu V_\nu \nabla^\mu V^\nu \right) + 12\, V^4,\nn \\ [2mm]
R^\pm_{\mu\nu}R^\pm{}^{\mu\nu} \!\!&=&\!\! R_{\mu\nu}R^{\mu\nu} - 4 \left( R_{\mu\nu} V^\mu V^\nu - RV^2  + \nabla_{[\mu}V_{\nu]} \nabla^\mu V^\nu\right) + 12\, V^4\,,\nn \\ [2mm]
(R^\pm)^2 \!\!&=&\!\! \left( R + 6\, V^2 \right)^2.
\eea
We see that
\be
R^-_{\rho\sigma\mu\nu} \ = \ R^+_{\mu\nu\rho\sigma}\,,\qquad R^-_{\nu\mu} \ = \ R^+_{\mu\nu}\,,
\ee 
Since the Riemann tensor is not symmetric under the exchange of the first and second pair of indices, and the Ricci tensor is not symmetric, two independent terms are
\bea
R^\pm_{\mu\nu\rho\sigma}R^\pm{}^{\rho\sigma\mu\nu} &=& R^-_{\mu\nu\rho\sigma}R^+{}^{\mu\nu\rho\sigma} \ = \  R^\pm_{\mu\nu\rho\sigma}R^\pm{}^{\mu\nu\rho\sigma} +  16\, \nabla_\mu V_\nu \nabla^\mu V^\nu\,, \nn \\ [2mm]
R^\pm_{\mu\nu}R^{\pm\,\nu\mu} &=& R^-_{\mu\nu} R^{+\mu\nu} \ = \ R^\pm_{\mu\nu} R^{\pm\mu\nu} +  8 \, \nabla_{[\mu} V_{\nu]} \nabla^\mu V^\nu  \,.
\eea
We will also need the relations
\bea
R^{\pm}_{\mu\nu \rho\sigma} *\!R^{\pm\,\mu\nu\rho\sigma} 
&=& R_{\mu\nu \rho\sigma} *\!R^{\,\mu\nu\rho\sigma} + 4i \nabla^\mu \lp \pm R V_\mu \mp 2 R_{\mu\nu}V^\nu \pm 2 V^2 \, V_\mu - \epsilon_{\mu\nu\rho\sigma} V^\nu \nabla^\rho V^\sigma  \rp  ,\nn \\ [2mm]
 R^\pm_{\mu\nu\rho\sigma} *\!R^\pm*^{\,\mu\nu\rho\sigma} &=& R^\pm_{\mu\nu\rho\sigma} R^\pm{}^{\rho\sigma\mu\nu} - 4 R^\pm_{\mu\nu} R^\pm{}^{\nu\mu} + (R^\pm)^2\nn \\ [2mm]
&=& R_{\mu\nu\rho\sigma} R^{\rho\sigma\mu\nu} - 4 R_{\mu\nu} R^{\nu\mu} + R^2 + 8\,\nabla_\nu \left( V_\mu \nabla^\mu V^\nu \right) \ .
\eea

\section{More on tensor calculus}\label{app:TensorCalculus}

\subsection{Multiplication rules}
\label{MultipletProduct}

In the following we recall the multiplet multiplication rules for two general multiplets $\scS_1, \scS_2$ with bottom components $C_1, C_2$ \cite{Sohnius:1982fw,Ferrara:1988qxa}, assuming that both the gravitino and its supersymmetry variations have been set to zero (see also \cite{Closset:2014uda}). These read
\bea
&& C \,=\, C_1 C_2~,\nn \\ [2mm] 
&& \chi \,=\, \chi_1 C_2 + (-1)^{F_1} \, C_1 \chi_2~,\qquad 
 \ti \chi \,=\,  \ti \chi_1 C_2 + (-1)^{F_1} \, C_1 \ti \chi_2~,\nn \\ [2mm] 
&& M \,=\, M_1 C_2 + C_1 M_2 - (-1)^{F_1} \, i \chi_1 \chi_2~,\qquad
 \ti M \,=\, \ti M_1 C_2 + C_1 \ti M_2 +  (-1)^{F_1} \, i \ti \chi_1 \ti \chi_2~,\nn \\ [2mm] 
&& a_\mu \,=\, a_{1 \mu} C_2 + C_1 a_{2 \mu} + (-1)^{F_1} \, \half \left(\chi_1 \sigma_\mu \ti \chi_2 - \ti \chi_1 \, \ti \sigma_\mu \chi_2\right)~,\nn \\ [2mm] 
&& \lambda \,=\, \lambda_1 C_2  + (-1)^{F_1} \,{i \over 2} \ti M_1 \chi_2 + {1 \over 2} \sigma^\mu \ti \chi_1 \left(a_{2\mu}-iD_\mu C_2\right)  \nn\\
&& \hskip20pt  +\, C_1 \lambda_2  + (-1)^{F_1} \,{i \over 2} \chi_1 \ti M_2 + {1 \over 2} \left(a_{1\mu}-iD_\mu C_1\right) \sigma^\mu \ti \chi_2 ~,\nn \\ [2mm] 
&& \ti \lambda \,=\, \ti \lambda_1 C_2  - (-1)^{F_1} \, {i \over 2} M_1 \ti \chi_2- {1 \over 2} \ti \sigma^\mu \chi_1 \left(a_{2\mu} + i D_\mu C_2\right) \nn\\
&&  \hskip20pt  +\, C_1 \ti \lambda_2  - (-1)^{F_1} \, {i \over 2} \ti \chi_1 M_2 - {1 \over 2} \left(a_{1\mu} + i D_\mu C_1 \right) \ti \sigma^\mu \chi_2 ~,\nn \\ [2mm] 
&& D \,=\, D_1 C_2 + C_1 D_2 + \half M_1 \ti M_2 + \half \ti M_1 M_2 - a_1^\mu a_{2 \mu} - D^\mu C_1 D_\mu C_2 \nn \\ [1mm] 
&& \hskip20pt - (-1)^{F_1} \chi_1 \Big(\lambda_2 +{i \over 2} \sigma^\mu D_\mu \ti \chi_2\Big) - (-1)^{F_1}  \ti \chi_1 \Big(\ti \lambda_2 +{i\over 2} \ti \sigma^\mu D_\mu \chi_2\Big) - (-1)^{F_1} \Big(\lambda_1 -{i \over 2} D_\mu \ti \chi_1 \ti \sigma^\mu\Big) \chi_2 \nn \\ [2mm] 
&& \hskip20pt  - (-1)^{F_1} \Big(\ti \lambda_1 -{i \over 2} D_\mu \chi_1 \sigma^\mu\Big) \ti \chi_2 + (-1)^{F_1}  {3 \over 2} V_\mu \left(\chi_1 \sigma^\mu \ti \chi_2 - \ti \chi_1 \ti \sigma^\mu \chi_2\right) \ .
\eea
Here $(-1)^{F_1}$ accounts for the Bose or Fermi statistics of $C_1$. The covariant derivative $D_\mu$ appearing here is defined as $D_\mu \, = \, \partial_\mu + \frac i2 \omega_\mu{}^{ab} S_{ab} - i\, r A_\mu \,.$

\subsection{Field strength multiplet and Yang--Mills Lagrangians}
\label{FieldStrength}

We now review the construction of field strength multiplets and associated Lagrangians.
Starting with a gauge vector multiplet in Wess--Zumino gauge $(a_{\mu},\lambda,\ti\lambda,D)$, one can construct the associated field strength multiplets $\Lambda_{\alpha}$ and $\ti \Lambda^{\dot\alpha}$. These are defined as the multiplets having $\lambda_\alpha$ and $\ti \lambda^{\dot\alpha}$ as bottom components, and turn out to be chiral and anti-chiral multiplets, respectively.
In components, at $\psi_{\mu}=\ti\psi_\mu=0$, they read
\bea
\Lambda_{\alpha} &=& \Big( \lambda_{\alpha}\,, \,  - \frac{1}{\sqrt 2} f_{ab} (\sigma^{ab})_{\alpha \beta} + \frac{i}{\sqrt 2}\, D\, \varepsilon_{\alpha \beta}\,, \, (i\,\sigma^{a} D^{-}_{a} \ti\lambda)_{\alpha} \Big)\ ,\nn  \\ [1mm]
\ti \Lambda^{\dot\alpha} &=& \Big( \ti\lambda^{\dot\alpha}\,, \, - \frac{1}{\sqrt 2} f_{ab} (\ti\sigma^{ab})^{\dot\alpha \dot\beta} -\frac{i}{\sqrt 2}\, D\, \varepsilon^{\dot\alpha \dot\beta}  \,, \,  (i\,\ti\sigma^{a} D^{-}_{a} \lambda)^{\dot\alpha} \Big)\ ,
\eea
where $f_{ab}$ is the field strength of $a_\mu$, and we have $\ti\sigma^{a} D^{-}_{a} \lambda = \ti\sigma^{a} (D_{a}\lambda + \frac {3i}{2} V_a \lambda)$ and $\sigma^{a} D^{-}_{a} \ti\lambda = \sigma^{a} (D_a \ti\lambda - \frac {3i}{2} V_a \ti\lambda)\,$.
Using the multiplet multiplication rules, we can then construct the chiral multiplet Tr$\,\Lambda_{\alpha} \Lambda^{\alpha}$, having R-charge $r= 2$, and the anti-chiral multiplet Tr$\,\ti \Lambda^{\dot\alpha} \ti \Lambda_{\dot\alpha}\,$,  having R-charge $r=-2$. The respective $F$-term and $\ti F$-term Lagrangians are the self-dual and anti-self-dual parts of the Yang--Mills Lagrangians, which again at vanishing gravitino read
\bea
\scL_{F} \big[{\rm Tr} \,   \Lambda^{\alpha} \Lambda_{\alpha}\big]
&=&   {\rm Tr} \, \Big[ \frac{1}{2} f_{ab}f^{ab} + \frac 12 f_{ab}*\!f^{ab} - D^2 + 2i\, \lambda \sigma^{a} D^{-}_{a} \ti\lambda  \, \Big]\ , \nn \\ [2mm]
\scL_{\ti F} \big[{\rm Tr} \, \ti \Lambda_{\dot\alpha} \ti \Lambda^{\dot\alpha}\big]
&=&  {\rm Tr} \, \Big[ \,  \frac{1}{2} f_{ab}f^{ab} - \frac 12 f_{ab} *\!f^{ab} - D^2 + 2i\, \ti\lambda\, \ti\sigma^{a} D^{-}_{a} \lambda \, \Big]\ . \label{SquareFieldStrMult}
\eea

Similarly we can construct Lagrangians mixing field strength multiplets of two different gauge vector multiplets:
\bea\label{MixedLagrangians}
\scL_F \big[ {\rm Tr}\, \Lambda_{1}{}^{\alpha}\Lambda_{2 \, \alpha} \big]
\!&=&\!  {\rm Tr} \, \Big[ \, \half f_1^{ab} f_{2 \, ab} + \frac 12 f_{1\,ab} *\!f_{2}^{ab} - D_1 D_2 + i\, \lambda_1 \sigma^{a} D^{-}_{a} \ti\lambda_2 + i\, \lambda_2 \sigma^{a} D^{-}_{a} \ti\lambda_1  \, \Big]\ , \nn\\ [2mm]
\scL_{\ti F} \big[{\rm Tr} \,\ti \Lambda_{1 \, \dot\alpha} \ti \Lambda^{2 \, \dot\alpha}\big]
\!&=&\!  {\rm Tr} \, \Big[ \, \half f_1^{ab} f_{2 \, ab} -\frac 12 f_{1\,ab} *\!f_{2}^{ab}  - D_1 D_2 + i\, \ti \lambda_1 \ti\sigma^{a} D^{-}_{a} \lambda_2 + i\, \ti\lambda_2 \ti\sigma^{a} D^{-}_{a} \lambda_1  \, \Big]\ . \nn\\
\eea
Note that $\mathcal L_F - \mathcal L_{\ti F}$ is a topological term in the field strengths.
This construction can be generalised with further vector multiplets and a kinetic matrix depending on chiral multiplets in a standard way.

\section{Comments on old minimal supergravity}\label{app:CompareOldMinimal}

It is interesting to compare the general analysis of the new minimal supergravity counterterms of section \ref{sec:VanishingResults} with a similar analysis in old minimal supergravity, pointing out the differences of the two approaches.
We will limit ourselves to a comparison of the vanishing arguments for $F$- and $\ti F$-terms. 

The supergravity fields in Euclidean old minimal supergravity are $(e^a_\mu, \psi_{\mu}, \ti\psi_\mu, b_\mu, M, \ti M)$, where, apart from the usual vierbein and gravitino, we have an auxiliary complex vector $b_\mu$ and two independent auxiliary complex scalars $M, \ti M$. 
Supersymmetric backgrounds are characterised by the vanishing of the gravitino supersymmetry variation:
\bea
 0 &=& \delta \psi_\mu \ = \ \nabla_\mu \zeta - \frac i6 M \sigma_\mu \ti\zeta - \frac i3 b_\mu \zeta - \frac i3 b^\nu \sigma_{\mu\nu} \zeta\ , \nn\\ [2mm]
  0 &=& \delta \ti\psi_\mu \ = \ \nabla_\mu \ti\zeta - \frac i6 \ti M \ti \sigma_\mu \zeta + \frac i3 b_\mu \ti \zeta + \frac i3 b^\nu \ti\sigma_{\mu\nu} \ti\zeta \ .
\eea
Solutions to these equations are given by pairs of spinors $(\zeta, \ti \zeta)$. The backgrounds that admit solutions have been discussed in \cite{Festuccia:2011ws} and classified in \cite{Dumitrescu:2012at}.

The chiral and anti-chiral multiplets have the same field content as in new minimal supergravity. The supersymmetry transformations on fermions at vanishing gravitino are the same as in new minimal supergravity. Their vanishing on a supersymmetric background is expressed by 
\bea
 0 &=& \delta \psi \ = \ \sqrt 2 \, \zeta F + i \sqrt 2 \, \sigma_\mu \ti\zeta \, \partial_\mu \phi\ , \nn\\ [2mm]
 0 &=& \delta \ti\psi \ = \ \sqrt 2 \, \ti\zeta \ti F + i \sqrt 2 \, \ti\sigma_\mu \zeta \, \partial_\mu \ti \phi \ .
 \label{OldMinConstraints}
\eea
We see that different cases must be considered. 

\begin{itemize}
\item In the presence of a single solution $(0, \ti \zeta)$, supersymmetric backgrounds are charaterized by $\ti F= 0$ and $F$ arbitrary. This is the same as in the new minimal supergravity approach when there is a single Killing spinor $\ti \zeta$ ;

\item In the presence of a pair of solutions $(0, \ti \zeta)$ and $(\zeta, 0)$, supersymmetric backgrounds are charaterized by $\ti F= F= 0$. This is the same as in the new minimal supergravity approach when there are two Killing spinors of opposite R-charge $\zeta, \ti \zeta$. Then all counterterms constructed as $F$- or $\ti F$-terms vanish in this case.

\item When the solutions are of the form $(\zeta, \ti \zeta)$ with $\zeta \neq 0$ and $\ti\zeta \neq 0$ almost everywhere (they can vanish on a subspace of zero measure), then the $F$- and $\ti F$-terms obey some constraints imposed by the equations \eqref{OldMinConstraints}, but they do not vanish in general. A prominent example is the supersymmetric background on $S^4$ \cite{Festuccia:2011ws} which admits four solutions of this type. Despite the large amount of supersymmetry preserved, it was shown in \cite{Gerchkovitz:2014gta} that a non-vanishing counterterm can be constructed as an $F$-term, leading to an ambiguity in the partition function of $\scN=1$ SCFTs on $S^4$.

\end{itemize}

The last point makes clear that there are cases where the ambiguities in the partition function of $\scN=1$ theories defined on a four-manifold using the rigid limit of old minimal and new minimal supergravities are different.


\bibliographystyle{JHEP}
\bibliography{Newbib}

\end{document}